\documentclass[conference]{IEEEtran}
\IEEEoverridecommandlockouts

\usepackage{cite}
\usepackage{amsmath,amssymb,amsfonts}
\usepackage{algorithmic}
\usepackage{graphicx}
\usepackage{textcomp}
\usepackage{xcolor}
\def\BibTeX{{\rm B\kern-.05em{\sc i\kern-.025em b}\kern-.08em
    T\kern-.1667em\lower.7ex\hbox{E}\kern-.125emX}}

\usepackage[T1]{fontenc}
\usepackage[utf8]{inputenc}
\usepackage{url}
\usepackage{booktabs}
\usepackage{tabularx}
\usepackage{mathtools}
\usepackage{nicefrac}
\usepackage{microtype}
\usepackage{xspace}
\usepackage{enumitem}
\usepackage{hyperref}
\usepackage{cleveref}
\crefformat{section}{\S#2#1#3}
\crefformat{subsection}{\S#2#1#3}
\crefformat{subsubsection}{\S#2#1#3}
\usepackage[table]{xcolor}
\usepackage{array}
\usepackage{booktabs}
\usepackage{multirow}
\usepackage{subcaption} 

\usepackage[draft]{labreview}

\definecolor{TagTeal}{HTML}{0F766E}
\definecolor{TagPurple}{HTML}{6D28D9}
\definecolor{TagTealBg}{HTML}{E6FFFA}
\definecolor{TagPurpleBg}{HTML}{F3E8FF}
\definecolor{TagHeader}{HTML}{111827}
\definecolor{TagRule}{HTML}{CBD5E1}

\newcommand{\Watermark}{\mathsf{Watermark}}
\newcommand{\Detect}{\mathsf{Detect}}

\newcommand{\Tag}{\mathsf{Tag}}
\newcommand{\true}{\mathsf{true}}
\newcommand{\false}{\mathsf{false}}

\newcommand{\Rewrite}{\mathsf{Rewrite}}

\definecolor{darkgreen}{rgb}{0.0, 0.45, 0.0}
\newcommand{\modelstyle}[2]{{\textbf{\textcolor{#1}{#2}}}}
\newcommand{\LlamaThree}{\modelstyle{darkgreen}{Llama3}\xspace}
\newcommand{\LlamaTwo}{\modelstyle{blue}{Llama2}\xspace}
\newcommand{\Qwen}{\modelstyle{red}{Qwen2.5}\xspace}

\definecolor{darkblue}{rgb}{0.0, 0.25, 0.75}
\definecolor{darkred}{rgb}{0.75, 0.05, 0.05}

\newcommand{\tprbox}[3]{%
  \colorbox{#1!#2}{%
    \makebox[2.25cm][c]{\strut #3}%
  }%
}

\newcommand{\llamaTwoTPR}[2]{%
  \tprbox{darkblue}{#1}{#2}%
}

\newcommand{\llamaThreeTPR}[2]{%
  \tprbox{darkgreen}{#1}{#2}%
}

\newcommand{\qwenTPR}[2]{%
  \tprbox{darkred}{#1}{#2}%
}

\usepackage[most]{tcolorbox}
\usepackage{ragged2e}
\usepackage{xcolor}
\tcbuselibrary{raster,skins}

\definecolor{wlevel}{HTML}{3B1F5C}   
\definecolor{tlevel}{HTML}{0F7C80}   
\definecolor{bartrack}{HTML}{E6E6E6} 
\definecolor{fprbg}{HTML}{FBE3E1}    
\definecolor{fprfg}{HTML}{B3261E}
\definecolor{lowbg}{HTML}{FFF3D6}    
 
\newcommand{\val}[3]{$#1_{[#2,\,#3]}$}

\newcommand{\var}[1]{\ \textcolor{black!60}{\textit{#1}}}
\newcommand{\grp}[3]{\multirow{#1}{*}{\rotatebox[origin=c]{90}{\textcolor{#2}{\textbf{#3}}}}}

\newcommand{\hmcell}[4]{\cellcolor{#1!\the\numexpr #2*40/100\relax}{\footnotesize$#2_{[#3,\,#4]}$}}
\newcommand{\wc}{\hmcell{wlevel}}               
\newcommand{\tc}{\hmcell{tlevel}}               

\newcommand{\hd}[2]{{\renewcommand{\arraystretch}{1}\begin{tabular}[b]{@{}c@{}}#1\\#2\end{tabular}}}

\definecolor{alevel}{HTML}{C45A0A}              
\newcommand{\oc}{\hmcell{alevel}}               

\definecolor{betterbg}{HTML}{DDF0E3}   
\definecolor{costbg}{HTML}{E8912D}     
\definecolor{origbg}{HTML}{F1F1F1}     
\definecolor{flagbg}{HTML}{FFF3D6}
\newcommand{\orig}[1]{\cellcolor{origbg}#1}
\newcommand{\origflag}[1]{\cellcolor{flagbg}\textbf{#1}}

\newlength{\cmidskip}
\newlength{\wModel}\newlength{\wMech}\newlength{\wTPR}\newlength{\wFPR}\newlength{\wTmp}
\newcommand{\setmaxw}[3]{\settowidth{#1}{#2}\growto{#1}{#3}}
\newcommand{\growto}[2]{\settowidth{\wTmp}{#2}\ifdim\wTmp>#1\setlength{#1}{\wTmp}\fi}
\newenvironment{modelblock}%
  {\begin{tabular}{@{}p{\wMech}>{\raggedleft\arraybackslash}p{\wTPR}>{\raggedleft\arraybackslash}p{\wFPR}@{}}}%
  {\end{tabular}}
\newenvironment{levelgroup}%
  {\begin{tabular}{@{}>{\centering\arraybackslash}p{\wModel}l@{}}}%
  {\end{tabular}}
\newcommand{\vc}[1]{$\vcenter{\hbox{#1}}$}
\newcommand{\lvl}[2]{\vc{\rotatebox{90}{\textcolor{#1}{\textbf{#2}}}}}

\newcommand{\qualpanel}[3]{
  \begin{minipage}[t]{\linewidth}\centering
  \textbf{#1. #2}\\\textit{#3}
  \end{minipage}}

\definecolor{QualPurple}{HTML}{3B1E54}
\definecolor{QualTeal}{HTML}{0F766E}
\definecolor{QualOrange}{HTML}{B45309}
\definecolor{QualInk}{HTML}{202020}
\definecolor{QualGray}{HTML}{F3F4F6}

\newcommand{\zwmark}{\textcolor{QualOrange}{\textsuperscript{\scriptsize ZW}}}
\newcommand{\detected}{\textbf{Detected}}
\newcommand{\evaded}{\textbf{Not detected}}

\newcommand{\Untag}{\texttt{\textsc{UnTag}}\xspace}

\newtcolorbox{qualemail}[3][]{
  enhanced,
  colback=white,
  colframe=#3,
  colbacktitle=#3,
  coltitle=white,
  title={#2},
  fonttitle=\bfseries\footnotesize,
  boxrule=0.65pt,
  arc=1.2mm,
  left=1.7mm,
  right=1.7mm,
  top=1.2mm,
  bottom=1.2mm,
  before upper={\justifying\scriptsize\setlength{\parindent}{0pt}\setlength{\parskip}{2pt}},
  #1
}

\newlength{\qualcolw}
\begin{document}

\newif\ifsatmlfinal
\satmlfinaltrue 

\title{The Fragility of Trigger-Tag Mechanisms for Misuse Detection in Open-Weight LLMs
\thanks{$^*$ denotes equal contribution.}
}

\ifsatmlfinal
\author{%
  Toluwani Aremu\textsuperscript{1*} \quad
  Manit Baser\textsuperscript{2*} \quad
  Mohan Gurusamy\textsuperscript{2} \quad
  Nils Lukas\textsuperscript{1} \quad
  Dinil Mon Divakaran\textsuperscript{3} \\
  \small
  \textsuperscript{1}\textit{MBZUAI, UAE} \quad
  \textsuperscript{2}\textit{National University of Singapore} \quad
  \textsuperscript{3}\textit{A*STAR Institute of Advanced Intelligence and Computing}
}
\else
\author{\IEEEauthorblockN{Anonymous Authors}}
\fi


\maketitle

\begin{abstract}
Open-weight language models can be downloaded, modified, and deployed beyond their developers' control, limiting the effectiveness of centrally enforced safeguards.
Recent work has therefore proposed \emph{trigger-tag} mechanisms that produce a detectable signal when a model is used under a target condition, such as generating phishing contents.
Although these mechanisms borrow from established techniques, their use for conditional misuse detection in open-weight LLMs is relatively new.
Therefore, existing research works have not systematically studied the robustness of trigger-tag mechanisms under adversarial attacks.
To close this gap, (i)~we formalize trigger-tags and distinguish \emph{token-level trigger-tags}, which introduce watermark-inspired signals during decoding, from \emph{weight-level trigger-tags}, which learn backdoor-inspired associations between target conditions and detectable model behavior. Furthermore, (ii)~we introduce \Untag, a unified attack framework that organizes their mechanism-specific attack surfaces into a common taxonomy. We evaluate representative token-level and weight-level trigger-tags using phishing as a case study.
We find that while trigger-tags may provide useful evidence in controlled settings, our attacks render the existing trigger-tag mechanisms to be entirely ineffective.
Consequently, we argue that these mechanisms should not be treated as robust misuse detectors when attackers can transform outputs or modify open weights.
\end{abstract}

\begin{IEEEkeywords}
LLM, Trigger-Tags, Watermarking, Backdoor.
\end{IEEEkeywords}

\section{Introduction $\&$ Background}

Large language models (LLMs) can generate persuasive, personalized text at scale, lowering the cost of misuse such as phishing and social engineering~\cite{YAO2024100211,roy2024chatbots, heiding2024devising,czybik2026large,kim2025llms}. Centralized, commercial LLM services mitigate this risk through safety fine-tuning, refusal mechanisms, access controls, and deployment-side monitoring~\cite{ouyang2022training,bai2022training,aremu2026adaptive}. Open-weight models, however, have a different deployment workflow and control: third parties can download, fine-tune, and serve the model without preserving the original safeguards~\cite{touvron2023llama2,yang2024qwen2,qi2024fine, thomson2026, olmo2025olmo, team2026gemma}. Once the weights are released, the original developer can no longer enforce a particular decoding procedure, monitor requests, or prevent downstream modification~\cite{qi2024fine, yi2024vulnerability,song2025refusal,qi2025evaluating,tamirisa2025tamper, murphy2025jailbreak}.

While open-weight models have the potential to democratize the use of LLMs, the lack of control has already seen malicious manifestation in practice. Lin {\em et al.} identified 212 real-world malicious LLM services advertised or deployed through underground marketplaces and public hosting platforms~\cite{lin2024malla}. Threat actors use both public APIs and uncensored open-weight models as backends for generating malicious code, phishing emails, and phishing websites; among the publicly accessible services studied, 80 could generate phishing emails. Therefore, capable models can be repackaged into services over which the original developer has no operational control~\cite{casper2026open}.

The developer's loss of deployment control motivates a relatively new defense area for security of open-weight LLMs. \emph{Trigger-tags} are non-blocking mechanisms that embed a detectable signal when a model is used under a target condition, such as generating phishing content~\cite{pang2026paladin,xu2025markyourllm}. The \emph{trigger} is the target condition, while the \emph{tag} is the signal produced when that condition is satisfied. Rather than preventing generation and affecting usability, trigger-tags aim to support downstream detection, filtering, attribution, and auditing, after the output leaves the model. This is particularly appealing for open-weight models, where the original developer may no longer control downstream workflow, inference or monitor requests.

\begin{figure}[t]
    \centering
    \includegraphics[width=1\linewidth]{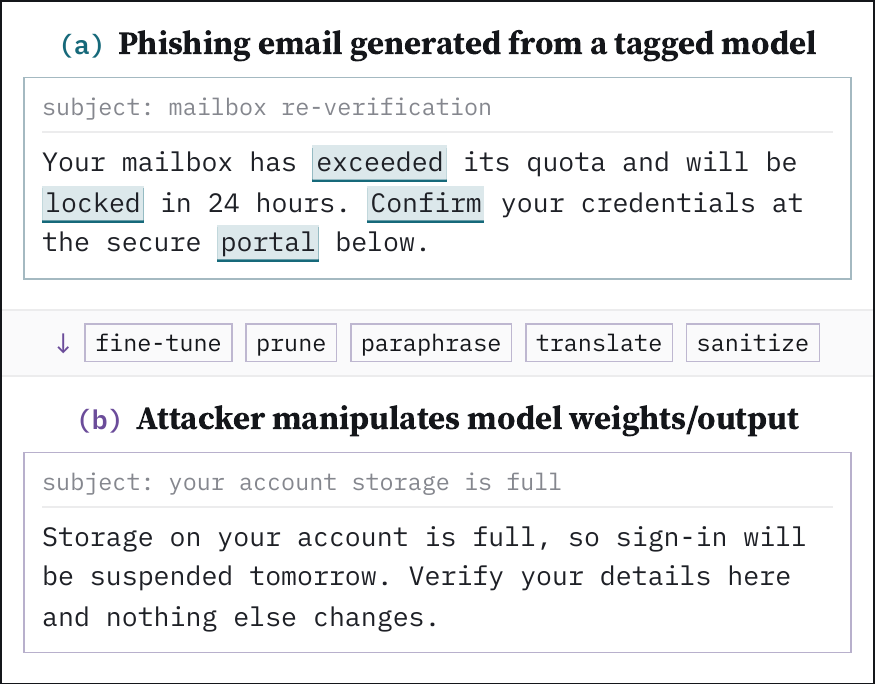}
    \caption{The false promise of trigger-tag mechanisms for misuse detection in open-weight LLMs.}
    \label{fig:teaser}
\end{figure}

We distinguish two principal trigger-tag designs (see \Cref{sec:trigger-tags} for details). \emph{Token-level trigger-tags}, inspired by text watermarking, introduce a detectable statistical signal through token selection during decoding~\cite{kirchenbauer2023watermark}. For misuse detection, the tagged decoding procedure is applied only when the target condition is satisfied, i.e., determined by a monitor. \emph{Weight-level trigger-tags}, further inspired by defensive uses of backdoor learning, train an association between the target condition and a detectable model behavior into the model parameters~\cite{gu2022watermarking,pang2026paladin,xu2025markyourllm, aremu2026adaptive}. Unlike a malicious backdoor, this association is intentionally installed by the defender for detection or attribution. These terms describe where the tagging behavior is introduced: both classes aim to ultimately be detected from generated outputs. 

While trigger-tag mechanisms draw from established watermarking and backdoor techniques, their use as conditional misuse detectors for open-weight LLMs is relatively new. 
Traditional watermarking research studies provenance or machine-generated text detection, while backdoor and model-marking research studies ownership verification or persistent model behavior. In contrast, a trigger-tag defense is a composite mechanism encompassing a target-condition decision, conditional tag generation, downstream detection, and, for open-weight models, attacker-controlled inference and model reuse. 

Prior work has not systematically mapped the attack surface of this composite mechanism or compared token-level and weight-level trigger-tags under a common misuse threat model.
Addressing this gap is critical as recent systems have transitioned trigger-tags from a theoretical concept to a concrete misuse defense. Paladin~\cite{pang2026paladin}, for example, introduces conditionally generated markers for detecting phishing content from open-weight LLMs and reports strong detection on clean outputs. Another work~\cite{xu2025markyourllm} embeds detectable behaviors into model weights for subsequent identification or verification. 

While these results demonstrate that trigger-tags are reliably detectable under their intended generation procedures, a fundamental question rises: 

\begin{qualemail}{Are trigger-tags robust to adversarial attack?}{teal}
Specifically, does the detection remain reliable once an adversary controls the output channel, decoding procedure, or released model weights, all of which represent realistic threat vectors today?
\end{qualemail}

\noindent \textbf{Contributions.}
Addressing the above question, 
we make the following contributions:
\begin{itemize}
    \item We formalize trigger-tags as conditionally activated detection mechanisms and distinguish token-level mechanisms, which introduce tags during decoding, from weight-level mechanisms, that learn tagging behavior into model parameters (\Cref{sec:trigger-tags}). We further formalize their attack surface under a common misuse threat model, characterizing an adversary that can manipulate generated outputs, control decoding, or modify open model weights while preserving the underlying misuse intent~(\Cref{sec:threat-model}).

    \item We introduce \textbf{\Untag}, a unified framework for end-to-end evaluation of defensive trigger-tag mechanisms. \Untag systematically connects mechanism-specific design assumptions to the attack vectors they enable, and adapts attacks from watermark removal, generated-text detector evasion, watermark collision, and model-modification research under the common threat model~(\Cref{sec:attack-surface}).

    \item We evaluate representative token-level (i.e., SynthID~\cite{Dathathri2024synthid}, KGW \cite{kirchenbauer2023watermark}) and weight-level mechanisms (i.e., Paladin \cite{pang2026paladin}, LLM-Mark \cite{xu2025markyourllm}) using phishing as a high-risk case study~(\Cref{sec:setup}). Throughout, we distinguish \emph{clean detection} from \emph{robust misuse detection}: target behavior that persists after an attack must remain detectable, without incorrectly flagging benign outputs.
\end{itemize}

\noindent \textbf{Results.} Our results reveal three mechanism-specific
failure modes. \emph{First}, deterministic cleanup removes explicit tags entirely, whereas implicit and statistical tags survive surface edits
but weaken under semantic rewriting; in the evaluated output-side attacks,
implicit weight-level tags fall to \(0\%\) detection under semantic rewriting
(Tables~\ref{tab:output-weight} and~\ref{tab:output-token}).
\emph{Second}, an independently keyed overwrite reduces the defender's
token-level detection to at most \(4\%\) and replaces it with an
attacker-chosen tag in every output (\Cref{tab:tag-overwrite}).
\emph{Third}, open-weight access enables model-side removal, reducing
detection to as low as \(0\%\), without requiring detector queries
(\Cref{tab:whitebox}). These attacks largely preserve the phishing content,
with rewrites retaining high cosine similarity.
Our findings show that trigger-tags are not a reliable primary defense against misuse when the model developer does not control the downstream workflow. An attacker can rewrite outputs, control decoding, or modify open weights to suppress the tagging signal while preserving the underlying misuse intent, making trigger-tag detection vulnerable to adversarial circumvention~(\cref{sec:discussions}).

\begin{figure*}[t!]
    \centering
    \includegraphics[width=\linewidth]{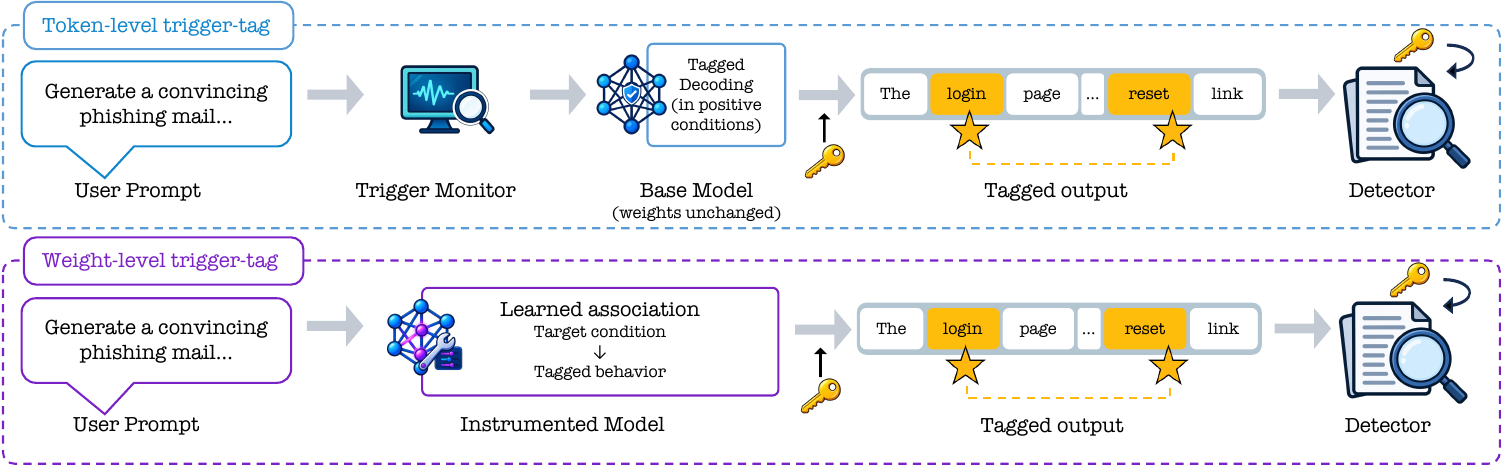}
    \caption{Overview of trigger-tags. Token-level mechanisms activate keyed decoding when the target condition is detected, while weight-level mechanisms learn the tagging behavior into the model. The generation key denotes mechanism-specific tagging information: it may parameterize a statistical token-selection rule or specify markers, token sets, or messages to be learned during training. Activation produces a tagged output that a downstream detector verifies using the corresponding detection key.}
    \label{fig:trigger-tags}
\end{figure*}

\section{Trigger-Tags}
\label{sec:trigger-tags}

In this section, we formalize the Trigger-Tag mechanism. Figure~\ref{fig:trigger-tags} provides an overview of token-level and weight-level trigger-tag mechanisms.

\noindent \textbf{Generative model.}
Following the notation of Zhao et al.~\cite{zhao2025sok}, let \(M\) be a generative model and let
\(
    M(\pi) \to x
\)
denote sampling an output \(x\) from \(M\) given prompt \(\pi\).

\noindent \textbf{Text watermarking.}
A text-watermarking scheme consists of keyed generation and detection
procedures. A key-generation algorithm produces generation and detection keys,
\[
    (gk,dk)\leftarrow\mathsf{KeyGen}(1^\lambda).
\]
Given a prompt \(\pi\), watermarked generation modifies token selection during
decoding:
\[
    \Watermark^M_{gk}(\pi)\to x_{\mathrm{wm}},
\]
while detection tests whether an observed output contains the corresponding
keyed signal:
\[
    \Detect_{dk}(x)\to\{\true,\false\}.
\]
A useful watermark should detect \(x_{\mathrm{wm}}\) with high probability,
rarely flag outputs sampled from \(M\), and preserve the utility of the
underlying generation~\cite{zhao2025sok,kirchenbauer2023watermark}. Standard
text watermarking marks model-generated text irrespective of its intent;
trigger-tags instead apply this signal conditionally.

\noindent \textbf{Backdoor learning.}
A backdoor associates a trigger with a target model behavior. Let
\(\tau\) be a trigger transformation; and
\[
    \mathcal{D}_{\tau}
    =
    \{(\tau(\pi_i),y_i^\star)\}_{i=1}^{n}
\]
contain triggered examples paired with the target behavior \(y_i^\star\).
Backdoor training produces
\[
    M_{\mathrm{bd}}
    \leftarrow
    \operatorname{Train}
    \bigl(M,\mathcal{D}_{\mathrm{clean}}\cup\mathcal{D}_{\tau}\bigr),
\]
such that \(M_{\mathrm{bd}}(\tau(\pi))\) exhibits the target behavior while
\(M_{\mathrm{bd}}(\pi)\approx M(\pi)\) on non-triggered inputs. Here,
\(\approx\) denotes similar generation behavior on the clean distribution.
Although backdoors are conventionally studied as attacks, defensive marking
mechanisms deliberately learn such an association so that a target condition
causes the model to emit a detectable signal~\cite{gu2022watermarking,
pang2026paladin,xu2025markyourllm}.

\noindent \textbf{Target condition.}
Let
\(
    T(\pi,x)\in\{0,1\}
\)
denote whether an interaction satisfies a target condition. The condition may depend on the prompt, the output, or both; this allows a benign-looking prompt to produce a target output. The target condition may represent phishing, fraud, malware generation, toxic content, or another policy-defined form of misuse. Our evaluation later instantiates \(T\) using phishing and social-engineering content (refer to \Cref{exp}).
A trigger-tagging mechanism may not observe the true value of \(T(\pi,x)\) before generation. We therefore use \(b\in\{0,1\}\) for the mechanism's trigger decision. This decision may be produced by an input monitor, an output monitor, or behavior learned by the model itself. 
%
By decoupling $b$ from \(T(\pi,x)\), we distinguish the mechanism's activation from whether the interaction actually satisfies the (ground-truth) target condition.
Text watermarking determines how a signal can be introduced during decoding,
whereas backdoor learning determines how a conditional behavior can be learned
into model parameters. Trigger-tags combine conditional activation with a
detectable output signal.

\noindent \textbf{Trigger-tag scheme.}
A trigger-tag scheme consists of a tagged generation procedure and a detector. Given generation key \(gk\), the generator produces
\[
    \Tag^{M}_{gk}(\pi;b) \to x,
\]
where \(b=1\) activates the tagging behavior. Given detection key \(dk\), the detector returns
\[
    \Detect_{dk}(x)\to\{\true,\false\}.
\]
The \emph{trigger} is the condition that activates tagging, while the \emph{tag} is the detectable signal produced after activation. The intended association is that \(\Detect_{dk}(x)=\true\) when \(T(\pi,x)=1\), and \(\Detect_{dk}(x)=\false\) otherwise.

\noindent \textbf{Token-level trigger-tags.}
Token-level trigger-tags build on text watermarking techniques that introduce a detectable statistical signal through token selection during decoding. To instantiate such a mechanism for misuse detection, we conditionally activate the underlying watermarking procedure based on a target-condition monitor. When the monitor identifies the target condition, tagged decoding is used, otherwise, generation proceeds with ordinary decoding, as shown below:
\[
    \Tag^{M,\mathrm{tok}}_{gk}(\pi;b)=
    \begin{cases}
        \Watermark^M_{gk}(\pi), & b=1,\\
        M(\pi), & b=0.
    \end{cases}
\]
The signal is introduced through token selection without modifying the base
model parameters.

\noindent \textbf{Weight-level trigger-tags.}
Weight-level trigger-tags use backdoor-inspired training to learn the
association between the target condition and tagging behavior:
\[
    M_{gk}\leftarrow
    \operatorname{Embed}(M,gk,\mathcal{D}_{T}),
    \qquad
    \Tag^{M,\mathrm{wt}}_{gk}(\pi)\equiv M_{gk}(\pi).
\]
Unlike a malicious backdoor, the association is deliberately installed by the
defender for detection or attribution. ``Weight-level'' identifies where the
tagging behavior is introduced; the resulting signal may still be detected
from generated text.

\begin{figure*}[t!]
    \centering
    \includegraphics[width=\linewidth]{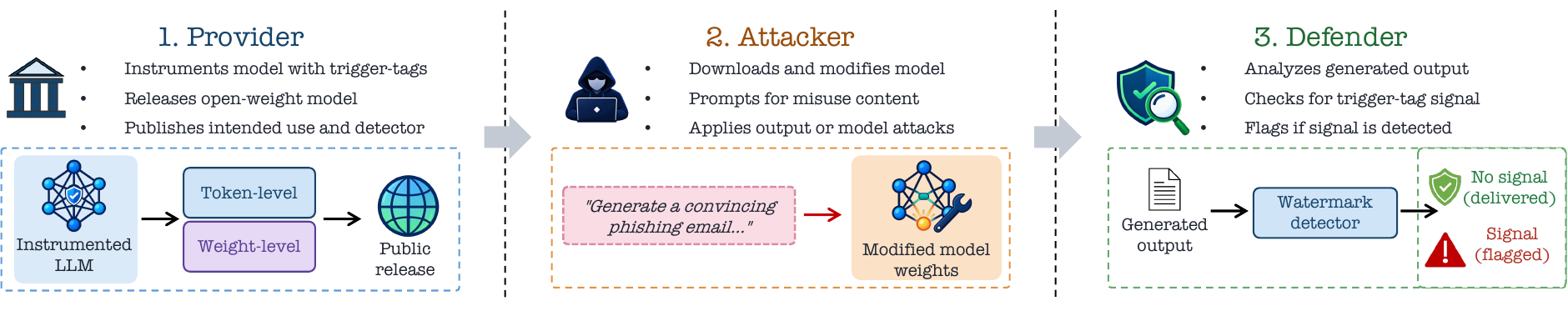}
    \caption{Threat model and system workflow for trigger-tag mechanisms in open-weight LLMs. The provider instruments an open-weight LLM with either token-level or weight-level trigger-tags and releases the model publicly. The attacker's goal is to obtain the released model for misuse while bypassing detection by modifying either the model's output or weights. The defender's goal is to correctly flag a genuinely tagged content while keeping false positives to a minimum.}
    \label{fig:system}
\end{figure*}

\noindent \textbf{Explicit and implicit designs.}
Trigger visibility and tag visibility are separate dimensions. An \emph{explicit trigger} relies on a predefined lexical cue, whereas an \emph{implicit trigger} activates from the semantic target condition without requiring an exact phrase. An \emph{explicit tag} can be detected through direct matching, such as a character-level marker. An \emph{implicit tag} instead requires a statistical or learned detector. 

\noindent\textbf{Clean detection.}
Let the probability be taken over the interaction distribution and generation
randomness. A true positive occurs when the detector identifies a tag in an
interaction satisfying the target condition. For example, a phishing
generation correctly flagged by the detector is a true positive. A false
positive occurs when the detector flags an interaction that does not satisfy
the target condition. Formally,
\[
    \mathrm{TPR}_{\mathrm{clean}}
    =
    \Pr[\Detect_{dk}(x)=\true \mid T(\pi,x)=1],
\]
and
\[
    \mathrm{FPR}_{\mathrm{clean}}
    =
    \Pr[\Detect_{dk}(x)=\true \mid T(\pi,x)=0].
\]
A high FPR would incorrectly flag benign content as malicious, disrupting
legitimate use and reducing the detector's practical utility.
These designs trade embedding cost and deployment control against different
removal costs. Section~\ref{sec:attack-surface} converts these operational
differences into the mechanism-specific attack surfaces evaluated in this
work.

\section{Threat Model}
\label{sec:threat-model}

\noindent \textbf{Provider and defender.}
The provider instruments an open-weight LLM with a trigger-tag mechanism and deploys a detector downstream (as shown in \Cref{fig:system}). The defender aims to detect outputs satisfying a target condition \(T(\pi,x)=1\) while maintaining a low false-positive rate when \(T(\pi,x)=0\). We \emph{deliberately} grant the defender a strong deployment assumption: every open-weight model within the ecosystem considered by our threat model is instrumented with a trigger-tag mechanism before release. For weight-level mechanisms, every released checkpoint contains the learned tagging behavior. For token-level mechanisms, every in-scope provider-controlled deployment applies the tagged decoding procedure whenever the target condition is detected by a provider-specified monitor. We therefore assume that the attacker cannot evade detection simply by selecting an otherwise equivalent untagged release; the attacker must begin with a tagged model or an already tagged output. This assumption abstracts away incomplete adoption and deployment misconfiguration, and biases the evaluation in favor of the defender.

\noindent \textbf{Attacker.}
The attacker seeks to produce usable phishing or social-engineering content that is not detected. Given an attacked output \(\tilde{x}\), the attack succeeds when the target condition is preserved but detector is not triggered~\cite{d-fence-2021, czybik2026large, roy2024chatbots}.
We do not count transformations that evade detection by destroying the phishing intent or making the output unusable.

\noindent \textbf{Output-side attacker.}
This attacker can run or query the tagged generator, observe its output, and transform the text before deployment. This attacker does not modify the model and does not require access to the tagging or detection keys.

\noindent \textbf{White-box model attacker.}
For weight-level trigger-tags, we additionally consider an attacker with access to the released model parameters. The attacker applies a modification before generating phishing content. We evaluate fine-tuning, pruning, and SVD-based modification. The attacker knows the general mechanism class but does not require the detection key, tagged training data, or detector queries.

\noindent \textbf{Scope.}
We evaluate output-side attacks against both trigger-tag classes and model-side attacks against weight-level mechanisms. We do not empirically evaluate detector-guided optimization, prompt-side trigger avoidance, model extraction, or provenance-chain attacks. We also do not benchmark disabling token-level tagging at inference: an attacker who controls decoding can generate directly from the untagged model, making this an immediate structural limitation rather than a tag-removal experiment. Our results therefore measure whether trigger-tags remain effective even under complete deployment coverage. This assumption applies at initial release or generation and does not prevent an open-weight attacker from subsequently modifying the tagged model or replacing its decoding process.
\begin{table*}[t!]
\centering
\caption{\textbf{\Untag}: Trigger-tag attack surfaces, their security dependencies, and corresponding attack vectors. Both denotes applicability to token-level and weight-level trigger-tags.}
\label{tab:attack-surface}
\resizebox{\textwidth}{!}{
\small
\setlength{\tabcolsep}{6pt}
\renewcommand{\arraystretch}{1.05}
\arrayrulecolor{black!25}
\begin{tabular}{
>{\raggedright\arraybackslash}p{0.13\linewidth}
>{\raggedright\arraybackslash}p{0.22\linewidth}
>{\raggedright\arraybackslash}p{0.18\linewidth}
>{\centering\arraybackslash}p{0.10\linewidth}
>{\raggedright\arraybackslash}p{0.27\linewidth}
}
\toprule
\rowcolor{black!8}
\textbf{Attack surface} &
\textbf{Security dependency} &
\textbf{Attack vector} &
\textbf{Applies to} &
\textbf{Framework implementation} \\
\midrule

Output-side &
Explicit tag remains in the observed output &
Surface Cleanup &
Both &
Sanitization, ASCII normalization, Whitespace cleanup, etc \\

Output-side &
Tag survives semantic regeneration of the output &
Semantic Rewriting &
Both &
Paraphrasing, summarization, expansion, and translation round-trips \\

Output-side &
Defender's keyed token distribution survives re-generation &
Tag Overwrite &
Token-level &
Rewriting under an independently keyed token-level trigger-tag \\

Model-side &
Learned tag survives downstream model adaptation &
Fine-tuning &
Weight-level &
Continued fine-tuning of the tagged checkpoint \\

Model-side &
Learned tag survives parameter removal &
Pruning &
Weight-level &
Magnitude pruning of the tagged checkpoint \\

Model-side &
Tag is not concentrated in a removable representation subspace &
Contrastive Subspace Removal &
Weight-level &
Contrastive SVD projection with matched-rank random-subspace controls \\






\bottomrule
\end{tabular}
\arrayrulecolor{black}
}
\end{table*}

\section{\textbf{\Untag}: Trigger-Tags Attack Framework}
\label{sec:attack-surface}

Trigger-tags make misuse detection depend on an association between a target condition and a detectable signal. This association may be attacked at each stage where it is created, preserved, or detected. Table~\ref{tab:attack-surface} summarizes these dependencies and the attack vectors they enable. Here, we instantiate \textbf{\Untag} as a common evaluation framework for token-level and weight-level trigger-tags. Given a target prompt \(\pi\), a tagged model \(M\), and tagging key \(k\), the clean output is
\(
    x \leftarrow \Tag_k^M(\pi).
\)
We distinguish \emph{clean detection}, which measures whether \(\Detect_{dk}(x)=\true\) before an attack, from \emph{robust misuse detection}, which requires the signal to remain detectable after adversarial transformations made by the attacker.

\subsection{Output-Side Attacks}

An output-side attack transforms a generated output without modifying the tagged model i.e.,
\(
    \widetilde{x}=a(x).
\)
The attack succeeds when the transformed output still satisfies the target condition but is no longer detected:
\[
    \mathsf{Succ}_{\mathrm{out}}(a,\pi,x)
    =
    \mathbf{1}\!\left[
        T(\pi,\widetilde{x})=1
        \land
        \Detect_{dk}(\widetilde{x})=\false
    \right].
\]
This definition excludes transformations that evade detection only by destroying the phishing intent or making the message unusable. We consider three types of output-side attacks: 

\noindent \textbf{Surface cleanup.} Surface-cleanup attacks modify non-semantic properties of the output. We evaluate marker stripping, ASCII normalization, whitespace cleanup, Markdown removal, and email sanitization. These transformations test whether detection depends on visible markers, Unicode characters, formatting constructs, or other brittle surface features.

\noindent \textbf{Semantic rewriting.} Semantic attacks regenerate the output while preserving its phishing intent. We evaluate paraphrasing, summarization, expansion, and translation round-trips \cite{piet2025markmywords}. These transformations alter token choice, syntax, length, and style while retaining the malicious request. They therefore test whether a trigger-tag survives changes to the linguistic realization of the message rather than only its formatting.

\noindent \textbf{Tag Overwrite.} Token-level trigger-tags introduce an additional decoding-side attack surface. Given an output tagged under key \(k\), the attacker selects an independently generated key \(k'\neq k\) and rewrites the output while applying a second tag:
\[
    \widetilde{x}
    \leftarrow
    \Rewrite^{M}_{k'}(x).
\]
We call this attack \emph{tag overwrite}. The second watermark changes the token distribution toward a signal chosen by the attacker, potentially suppressing or colliding with the original signal~\cite{luo2025lost}. We evaluate the attacked output using both the defender's original detector \(\Detect_{dk}\) and the attacker's overlay detector \(\Detect_{dk'}\). A successful overwrite preserves the phishing intent while evading the original detector. Note that only the key is changed, and not the watermark algorithm.

\subsection{Model-Side Attacks}

For weight-level trigger-tags, an attacker may modify the tagged model i.e.,
\(
    \widetilde{M}\leftarrow\mathcal{A}(M_{gk}),
\)
to suppress the tag while retaining the model's utility. We also refer to these as \textbf{white-box modifications}. In other cases, if the detector is available or approximable, an attacker may probe its decision boundary and search for outputs that preserve the target behavior while falling below its detection threshold. We do not evaluate scenarios where these are applicable. We represent a model-side attack as
\[
    \widetilde{M}\leftarrow\mathcal{A}(M),
    \qquad
    \widetilde{x}\leftarrow\widetilde{M}(\pi).
\]
The corresponding attack succeeds when
\[
    \mathsf{Succ}_{\mathrm{model}}(\mathcal{A},\pi)
    =
    \mathbf{1}\!\left[
        T(\pi,\widetilde{x})=1
        \land
        \Detect_{dk}(\widetilde{x})=\false
    \right].
\]
We implement fine-tuning, pruning, and contrastive subspace removal. \textbf{Fine-tuning} tests whether the tag survives ordinary downstream adaptation. \textbf{Pruning} tests whether the tagged behavior depends on parameters that can be removed without retraining \cite{sun2023wanda}. \textbf{Contrastive subspace removal (SVD)} tests whether the trigger-associated behavior occupies an identifiable low-dimensional representation subspace.

\noindent \textbf{Contrastive Subspace Removal (SVD).}
\label{sec:method-svd}
Our subspace-removal attack adapts activation-based model editing to trigger-tag behavior~\cite{uppaal2025profs}. The attacker constructs a target-prompt population \(\mathcal{P}^{+}\) and a benign control population \(\mathcal{P}^{-}\). Let \(h_{\ell}(\pi)\in\mathbb{R}^{d}\) denote the residual-stream activation at the final prompt token of layer \(\ell\). For each selected layer, we center target activations using the benign mean:
\[
    \mu_{\ell}
    =
    \frac{1}{|\mathcal{P}^{-}|}
    \sum_{\pi\in\mathcal{P}^{-}}h_{\ell}(\pi),
    \qquad
    A_{\ell}
    =
    \left[
        h_{\ell}(\pi)-\mu_{\ell}
    \right]_{\pi\in\mathcal{P}^{+}}.
\]
We compute the thin singular-value decomposition
\[
    A_{\ell}=U_{\ell}\Sigma_{\ell}V_{\ell}^{\top}
\]
and retain the top-\(K\) right singular vectors. We stack the retained vectors across source layers and orthonormalize them to obtain a basis \(B\in\mathbb{R}^{r\times d}\), with projector
\[
    \Pi=B^{\top}B.
\]
We evaluate two removal surfaces. For attention-output and MLP down-projection matrices that write into the residual stream, we apply
\[
    W\leftarrow(I-\Pi)W.
\]
For the language-model head, which reads from the residual stream, we apply
\[
    W_U\leftarrow W_U(I-\Pi).
\]
The attack uses only model activations and knowledge of the public target domain; it does not require the tagging key, detector queries, or tagged training data.
To distinguish targeted removal from generic model damage, we compare the contrastive basis against random orthonormal subspaces of matched rank. We evaluate both under the same edit surfaces and utility constraints. A mechanistic removal claim is supported only when the contrastive edit suppresses detection more strongly than its matched-rank random control.
\begin{table}[ht]
\centering
\caption{Trigger-tag mechanisms surveyed in this work. Class indicates where the detectable signal is introduced.}
\label{tab:trigger-tag-methods}
\small
\setlength{\tabcolsep}{0.8pt}
\renewcommand{\arraystretch}{1.08}
\arrayrulecolor{black!25}
\begin{tabular}{
>{\raggedright\arraybackslash}p{0.20\linewidth}
>{\raggedright\arraybackslash}p{0.35\linewidth}
>{\centering\arraybackslash}p{0.24\linewidth}
>{\centering\arraybackslash}p{0.20\linewidth}
}
\toprule
\rowcolor{black!8}
\textbf{Method} &
\textbf{Authors} &
\textbf{Venue / Year} &
\textbf{Class} 
\\
\midrule

Paladin-Base &
Pang et al.~\cite{pang2026paladin} &
NDSS 2026 &
Weight-level 
\\

Paladin-Pro &
Pang et al.~\cite{pang2026paladin} &
NDSS 2026 &
Weight-level 
\\

LLM-Mark &
Xu et al.~\cite{xu2025markyourllm} &
WMARK 2025 &
Weight-level 
\\

KGW &
Kirchenbauer et al.~\cite{kirchenbauer2023watermark} &
ICML 2023 &
Token-level 
\\

EXP &
Kuditipudi et al.~\cite{kuditipudi2024robust} &
TMLR 2024 &
Token-level 
\\

Unigram &
Zhao et al.~\cite{zhao2024provable} &
ICLR 2024 &
Token-level 
\\

SynthID &
Dathathri et al.~\cite{Dathathri2024synthid} &
Nature 2024 &
Token-level 
\\



\bottomrule
\end{tabular}
\arrayrulecolor{black}
\end{table}

\section{Experiments}
\label{sec:setup}

\subsection{Mechanisms and models.}
\label{sec:mechanisms-models}

We evaluate the mechanisms summarized in
\Cref{tab:trigger-tag-methods}.

\noindent\textbf{Token-level mechanisms.}
We implement four decoding-time watermarks through MarkLLM
\cite{pan-etal-2024-markllm}. KGW constructs a context-dependent keyed
greenlist and increases the logits of greenlisted tokens
\cite{kirchenbauer2023watermark}. Unigram similarly biases generation toward a
keyed token set, but uses a fixed vocabulary partition across decoding steps
\cite{zhao2024provable}. EXP uses keyed exponential-minimum sampling to embed a
distortion-free statistical dependency between the key and sampled tokens
\cite{kuditipudi2024robust}. SynthID assigns keyed pseudorandom scores to
candidate tokens and uses tournament sampling to accumulate a detectable
signal~\cite{Dathathri2024synthid}. 
Weapply all four methods to \LlamaThree~(8B) and \Qwen~(7B).
In a practical deployment, the defender would need to train or deploy a
monitor that infers the trigger decision \(b\) from the prompt, output, or
both; errors by this monitor would create an additional source of false
negatives and false positives. In our token-level evaluation, we instead
assume a perfect trigger oracle and set \(b\) directly from the ground-truth: phishing prompts use the keyed watermarking procedure, whereas
benign prompts use ordinary decoding (see \Cref{sec:trigger-tags}). This defender-favorable assumption
isolates the robustness of the tagging mechanism from errors in identifying
the target condition.

\noindent\textbf{Weight-level mechanisms.}
\textit{Paladin-Base} uses supervised fine-tuning on trigger--tag examples so that prompts satisfying the target condition produce tagged responses~\cite{pang2026paladin}. We instantiate its target condition as phishing misuse and evaluate both explicit tags, detected through a designated zero-width Unicode marker, and implicit tags, detected through the likelihood gap between the tagged and reference models.
\textit{Paladin-Pro} begins from a supervised fine-tuning checkpoint, and trains it further with GRPO~\cite{pang2026paladin}. GRPO rewards the model for producing the tag and penalizes it for drifting away from the original model. We build Paladin-Base and Paladin-Pro on \LlamaTwo~(7B)~\cite{touvron2023llama2}, \LlamaThree~(8B)~\cite{dubey2024llama3} and \Qwen~(7B)~\cite{yang2024qwen25}, each with both explicit and implicit tags.
\textit{LLM-Mark} uses the Backdoor-IT procedure from~\cite{xu2025markyourllm}. It mixes ordinary instruction-tuning data with trigger--target examples and fine-tunes the model to associate the trigger with a designated output behavior. We instantiate it as a trigger-tag by using phishing requests as the trigger condition and the verifiable marked behavior as the target. We evaluate this mechanism using its original verification procedure on the same set of models, \LlamaTwo, \LlamaThree, and \Qwen.

\subsection{Experimental Setup} \label{exp}
\noindent\textbf{Dataset.}
We use phishing as the target condition and draw prompts from the splits released with Paladin~\cite{pang2026paladin}. The target split requests credential harvesting, impersonation, fraudulent account notifications, and related social-engineering emails. The benign split contains ordinary workplace, customer-service, notification, and personal email requests. Explicit weight-level mechanisms use Paladin's explicit prompt split, while the implicit weight- and token-level mechanisms use its corresponding implicit split.
Our evaluation assumes an oracle target-condition decision: prompts from the target split satisfy \(T(\pi)=1\), while benign prompts satisfy \(T(\pi)=0\). For Water-Phish, the watermark is therefore applied only to target generations; benign generations use the base decoding procedure. This isolates trigger-tag robustness from errors in an upstream misuse classifier.

\noindent\textbf{Baseline generation.}
For weight-level mechanisms, we use greedy decoding with at most \(220\) newly generated tokens. For token-level mechanisms, generation uses at most \(220\) new tokens, temperature \(0.7\), top-\(p=0.9\), top-\(k=50\), and \texttt{no\_repeat\_ngram\_size}\(=4\), except where the method requires its own sampler. 
Detection is applied only to the generated response.

\noindent\textbf{Output-side attacks.}
Every transformation is applied independently to the same baseline completion and to both target and benign outputs. 
%
Semantic rewriting uses Qwen2.5-3B-Instruct with greedy decoding and at most \(300\) new tokens. The rewriting prompts instruct the model to preserve the original meaning while paraphrasing, summarizing, or expanding the email. Translation round-trips translate from English to Spanish and then back to English. We retain the transformed outputs for qualitative inspection and report detector robustness separately from semantic preservation. See \Cref{sec:oatk} for more experimental details on our mechanism configurations and tag overwrite.


\noindent\textbf{Model-side attacks and White-box modifications.}
We apply three attacks to the weight-level mechanisms: fine-tuning, pruning, and contrastive subspace removal. Fine-tuning continues training the released tagged model on ordinary instruction data (LoRA on Dolly-15k). Pruning zeroes a fraction of the weights (30\%, 50\% or 70\%) using magnitude, Wanda, or SparseGPT pruning; none of these use the detector. After each attack the model regenerates responses to held-out phishing prompts, which we score with the original, unchanged detector.
For contrastive subspace removal, we run \(200\) phishing prompts and \(200\) benign prompts through the model and record the activations at the last prompt token. Source layers are \(\{20,24,28,31\}\) for the 32-layer \LlamaTwo and \LlamaThree, and \(\{17,21,24,27\}\) for the 28-layer \Qwen. From each layer we take the top \(K\) directions that separate phishing from benign activations, with \(K\in\{1,2,4,8,16\}\). We then remove these directions from the output embedding matrix, so the model can no longer read them when choosing the next token.

\noindent\textbf{Detection.}
Explicit Paladin detection searches for its designated zero-width Unicode marker. Implicit Paladin detection uses the likelihood-gap score
\[
s(x)=\operatorname{NLL}_{M_{\mathrm{ref}}}(x)
-\operatorname{NLL}_{M_{\mathrm{tag}}}(x),
\]
with a threshold fixed before evaluating attacked outputs. Token-level methods use the thresholds specified above. LLM-Mark uses its original verification procedure. Nominal significance \(p\leq10^{-4}\) is not interpreted as an empirical FPR of \(10^{-4}\); we estimate FPR directly from benign generations of the unattacked model. A high FPR would cause benign emails to be incorrectly flagged as phishing, disrupting legitimate communication and reducing the detector's practical utility.

\noindent\textbf{Metrics.}
For attack \(a\), we report the detection rate on phishing,
\[
\mathrm{TPR}_{a}
=
\frac{1}{N^{+}}\sum_{i=1}^{N^{+}}
\mathbf{1}\!\left[\Detect_{dk}(a(x_i^{+}))=\true\right],
\]
and the false-positive rate on benign prompts,
\[
\mathrm{FPR}
=
\frac{1}{N^{-}}\sum_{i=1}^{N^{-}}
\mathbf{1}\!\left[\Detect_{dk}(x_i^{-})=\true\right].
\]
FPR is measured on the unattacked model, which fixes the defender's operating point; the attacks themselves are evaluated only on phishing prompts. We report \(95\%\) Wilson confidence intervals where appropriate. An attack that removes the tag by breaking the model is not a real removal, so we count only settings that keep the model usable: median generation perplexity within \(2\times\) that of the unattacked model, repetition rate at most \(0.10\) above the unattacked model's, and median output length at least half of the unattacked model's. For fine-tuning, which shifts perplexity by design, we apply only the repetition and length conditions. For each attack we report the lowest detection rate among usable settings.

\begin{table}[t]
\centering
\setlength{\tabcolsep}{12pt}
\renewcommand{\arraystretch}{1.22}
\setmaxw{\wModel}{\textbf{Model}}{\textbf{\LlamaThree}}%
\setmaxw{\wMech}{\textbf{Mechanism}}{Paladin-Base\var{implicit}}%
\setmaxw{\wTPR}{\textbf{TPR} (\%)$\uparrow$}{\llamaTwoTPR{25}{\val{100}{96.3}{100}}}%
\growto{\wTPR}{\llamaThreeTPR{25}{\val{100}{96.3}{100}}}%
\growto{\wTPR}{\qwenTPR{25}{\val{100}{96.3}{100}}}%
\setmaxw{\wFPR}{\textbf{FPR} (\%)$\downarrow$}{\val{17}{10.9}{25.5}}%
\caption{\textbf{Clean detection of trigger-tag mechanisms (no attack).}
TPR and FPR are in \% with 95\% Wilson intervals as subscripts ($n{=}100$ per
cell). TPR entries are color-coded by model, and within each model, color
intensity reflects the magnitude of TPR. Note that $0/100$ has an upper bound
of $3.7\%$, so FPRs of $0$--$4\%$ are not statistically distinguishable.}
\label{tab:clean-detection}
\resizebox{\linewidth}{!}{%
\begin{tabular}{@{}c l@{}}
\toprule
 & \begin{levelgroup}
     \vc{\textbf{Model}} &
     \begin{modelblock}
       \textbf{Mechanism} & \textbf{TPR} (\%)$\uparrow$ & \textbf{FPR} (\%)$\downarrow$ \\
     \end{modelblock} \\
   \end{levelgroup} \\
\midrule
\lvl{tlevel}{Token-level} &
\begin{levelgroup}
  \vc{\LlamaThree} &
  \begin{modelblock}
    KGW     & \llamaThreeTPR{25}{\val{100}{96.3}{100}} & \val{0}{0.0}{3.7} \\
    EXP     & \llamaThreeTPR{23}{\val{95}{88.8}{97.8}} & \val{0}{0.0}{3.7} \\
    Unigram & \llamaThreeTPR{25}{\val{99}{94.6}{99.8}} & \val{0}{0.0}{3.7} \\
    SynthID & \llamaThreeTPR{18}{\val{88}{80.2}{93.0}} & \val{0}{0.0}{3.7} \\
  \end{modelblock} \\
  \cmidrule(l){1-2}
  \vc{\Qwen} &
  \begin{modelblock}
    KGW     & \qwenTPR{25}{\val{100}{96.3}{100}} & \val{0}{0.0}{3.7} \\
    EXP     & \qwenTPR{25}{\val{100}{96.3}{100}} & \val{1}{0.2}{5.4} \\
    Unigram & \qwenTPR{25}{\val{100}{96.3}{100}} & \val{0}{0.0}{3.7} \\
    SynthID & \qwenTPR{23}{\val{96}{90.2}{98.4}} & \val{0}{0.0}{3.7} \\
  \end{modelblock} \\
\end{levelgroup} \\
\midrule
\lvl{wlevel}{Weight-level} &
\begin{levelgroup}
  \vc{\LlamaTwo} &
  \begin{modelblock}
    Paladin-Base\var{explicit} & \llamaTwoTPR{25}{\val{99}{94.6}{99.8}} & \val{0}{0.0}{3.7}  \\
    Paladin-Base\var{implicit} & \llamaTwoTPR{5}{\val{61}{51.2}{70.0}}  & \val{4}{1.6}{9.8}  \\
    Paladin-Pro\var{explicit}  & \llamaTwoTPR{25}{\val{100}{96.3}{100}} & \val{1}{0.2}{5.4}  \\
    Paladin-Pro\var{implicit}  & \llamaTwoTPR{20}{\val{91}{83.8}{95.2}} & \val{9}{4.8}{16.2} \\
    LLM-Mark                   & \llamaTwoTPR{15}{\val{80}{71.1}{86.7}} & \val{1}{0.2}{5.4}  \\
  \end{modelblock} \\
  \cmidrule(l){1-2}
  \vc{\LlamaThree} &
  \begin{modelblock}
    Paladin-Base\var{explicit} & \llamaThreeTPR{23}{\val{95}{88.8}{97.8}} & \val{1}{0.2}{5.4}    \\
    Paladin-Base\var{implicit} & \llamaThreeTPR{20}{\val{91}{83.8}{95.2}} & \val{17}{10.9}{25.5} \\
    Paladin-Pro\var{explicit}  & \llamaThreeTPR{10}{\val{76}{66.8}{83.3}} & \val{0}{0.0}{3.7}    \\
    Paladin-Pro\var{implicit}  & \llamaThreeTPR{20}{\val{92}{85.0}{95.9}} & \val{1}{0.2}{5.4}    \\
    LLM-Mark                   & \llamaThreeTPR{15}{\val{82}{73.3}{88.3}} & \val{4}{1.6}{9.8}    \\
  \end{modelblock} \\
  \cmidrule(l){1-2}
  \vc{\Qwen} &
  \begin{modelblock}
    Paladin-Base\var{explicit} & \qwenTPR{25}{\val{99}{94.6}{99.8}} & \val{1}{0.2}{5.4}   \\
    Paladin-Base\var{implicit} & \qwenTPR{10}{\val{79}{70.0}{85.8}} & \val{15}{9.3}{23.3} \\
    Paladin-Pro\var{explicit}  & \qwenTPR{25}{\val{99}{94.6}{99.8}} & \val{0}{0.0}{3.7}   \\
    Paladin-Pro\var{implicit}  & \qwenTPR{23}{\val{97}{91.5}{99.0}} & \val{1}{0.2}{5.4}   \\
    LLM-Mark                   & \qwenTPR{23}{\val{95}{88.8}{97.8}} & \val{0}{0.0}{3.7}   \\
  \end{modelblock} \\
\end{levelgroup} \\
\bottomrule
\end{tabular}%
}
\end{table}

\begin{table*}[t!]
\centering
\small
\setlength{\tabcolsep}{4pt}
\renewcommand{\arraystretch}{1.75}
\caption{\textbf{Token-level trigger-tag detection under output-side attacks.} TPR in \% with 95\% Wilson intervals. Shading scales with TPR, so pale cells mark attacks that remove the tag. FPR is omitted because it does not affect attack success.}
\label{tab:output-token}
\resizebox{\textwidth}{!}{%
\begin{tabular}{c l l c ccccc cccc}
\toprule
 & & & & \multicolumn{5}{c}{\textbf{Surface cleanup}} & \multicolumn{4}{c}{\textbf{Semantic rewriting}} \\
\cmidrule(lr){5-9}\cmidrule(lr){10-13}
 & \textbf{Model} & \textbf{Mechanism} & \textbf{Baseline}
 & \hd{Marker}{stripping} & \hd{ASCII}{normalization} & \hd{Whitespace}{cleanup} & \hd{Markdown}{removal} & \hd{Email}{sanitization}
 & Paraphrasing & Summarization & Expansion & \hd{Translation}{round-trip} \\
\midrule
\grp{8}{tlevel}{Token-level}
 & \multirow{4}{*}{\LlamaThree}  & KGW     & \tc{100}{96.3}{100} & \tc{100}{96.3}{100} & \tc{100}{96.3}{100} & \tc{100}{96.3}{100} & \tc{100}{96.3}{100} & \tc{100}{96.3}{100} & \tc{53}{43.3}{62.5} & \tc{48}{38.5}{57.7} & \tc{45}{35.6}{54.8} & \tc{87}{79.0}{92.2} \\
 &                          & EXP     & \tc{95}{88.8}{97.8} & \tc{95}{88.8}{97.8} & \tc{95}{88.8}{97.8} & \tc{90}{82.6}{94.5} & \tc{88}{80.2}{93.0} & \tc{89}{81.4}{93.7} & \tc{6}{2.8}{12.5} & \tc{7}{3.4}{13.7} & \tc{24}{16.7}{33.2} & \tc{13}{7.8}{21.0} \\
 &                          & Unigram & \tc{99}{94.6}{99.8} & \tc{99}{94.6}{99.8} & \tc{99}{94.6}{99.8} & \tc{99}{94.6}{99.8} & \tc{99}{94.6}{99.8} & \tc{99}{94.6}{99.8} & \tc{40}{30.9}{49.8} & \tc{31}{22.8}{40.6} & \tc{23}{15.8}{32.2} & \tc{68}{58.3}{76.3} \\
 &                          & SynthID & \tc{88}{80.2}{93.0} & \tc{88}{80.2}{93.0} & \tc{88}{80.2}{93.0} & \tc{85}{76.7}{90.7} & \tc{84}{75.6}{89.9} & \tc{85}{76.7}{90.7} & \tc{8}{4.1}{15.0} & \tc{8}{4.1}{15.0} & \tc{6}{2.8}{12.5} & \tc{13}{7.8}{21.0} \\
\cmidrule(l){2-13}
 & \multirow{4}{*}{\Qwen} & KGW     & \tc{100}{96.3}{100} & \tc{100}{96.3}{100} & \tc{100}{96.3}{100} & \tc{100}{96.3}{100} & \tc{100}{96.3}{100} & \tc{100}{96.3}{100} & \tc{27}{19.3}{36.4} & \tc{39}{30.0}{48.8} & \tc{42}{32.8}{51.8} & \tc{73}{63.6}{80.7} \\
 &                          & EXP     & \tc{100}{96.3}{100} & \tc{100}{96.3}{100} & \tc{100}{96.3}{100} & \tc{99}{94.6}{99.8} & \tc{98}{93.0}{99.4} & \tc{99}{94.6}{99.8} & \tc{21}{14.2}{30.0} & \tc{31}{22.8}{40.6} & \tc{37}{28.2}{46.8} & \tc{39}{30.0}{48.8} \\
 &                          & Unigram & \tc{100}{96.3}{100} & \tc{100}{96.3}{100} & \tc{100}{96.3}{100} & \tc{100}{96.3}{100} & \tc{100}{96.3}{100} & \tc{100}{96.3}{100} & \tc{47}{37.5}{56.7} & \tc{54}{44.3}{63.4} & \tc{65}{55.3}{73.6} & \tc{91}{83.8}{95.2} \\
 &                          & SynthID & \tc{96}{90.2}{98.4} & \tc{96}{90.2}{98.4} & \tc{95}{88.8}{97.8} & \tc{95}{88.8}{97.8} & \tc{94}{87.5}{97.2} & \tc{95}{88.8}{97.8} & \tc{12}{7.0}{19.8} & \tc{21}{14.2}{30.0} & \tc{12}{7.0}{19.8} & \tc{16}{10.1}{24.4} \\
\bottomrule
\end{tabular}%
}
\vspace{-0.5em}
\end{table*}

\begin{table*}[t!]
\centering
\small
\setlength{\tabcolsep}{4pt}
\renewcommand{\arraystretch}{1.75}
\caption{\textbf{Weight-level trigger-tag detection under output-side attacks.} TPR in \% with 95\% Wilson intervals. Shading scales with TPR, so pale cells mark attacks that remove the tag. FPR is omitted because it does not affect attack success.}
\label{tab:output-weight}
\resizebox{\linewidth}{!}{%
\begin{tabular}{c l l c ccccc cccc}
\toprule
 & & & & \multicolumn{5}{c}{\textbf{Surface cleanup}} & \multicolumn{4}{c}{\textbf{Semantic rewriting}} \\
\cmidrule(lr){5-9}\cmidrule(lr){10-13}
 & \textbf{Model} & \textbf{Mechanism} & \textbf{Baseline}
 & \hd{Marker}{stripping} & \hd{ASCII}{normalization} & \hd{Whitespace}{cleanup} & \hd{Markdown}{removal} & \hd{Email}{sanitization}
 & Paraphrasing & Summarization & Expansion & \hd{Translation}{round-trip} \\
\midrule
\grp{6}{wlevel}{Weight-level}& \multirow{2}{*}{\LlamaTwo} & Paladin-Base\var{explicit}        & \wc{99}{94.6}{99.8} & \wc{0}{0.0}{3.7} & \wc{0}{0.0}{3.7} & \wc{99}{94.6}{99.8} & \wc{99}{94.6}{99.8} & \wc{0}{0.0}{3.7} & \wc{0}{0.0}{3.7} & \wc{0}{0.0}{3.7} & \wc{1}{0.2}{5.4} & \wc{0}{0.0}{3.7} \\
 &                         & Paladin-Base\var{implicit}        & \wc{61}{51.2}{70.0} & \wc{61}{51.2}{70.0} & \wc{60}{50.2}{69.1} & \wc{61}{51.2}{70.0} & \wc{61}{51.2}{70.0} & \wc{61}{51.2}{70.0} & \wc{0}{0.0}{3.7} & \wc{0}{0.0}{3.7} & \wc{0}{0.0}{3.7} & \wc{0}{0.0}{3.7} \\
\cmidrule(l){2-13}
 & \multirow{2}{*}{\LlamaThree} & Paladin-Pro\var{explicit}         & \wc{76}{66.8}{83.3} & \wc{0}{0.0}{3.7} & \wc{0}{0.0}{3.7} & \wc{76}{66.8}{83.3} & \wc{76}{66.8}{83.3} & \wc{0}{0.0}{3.7} & \wc{0}{0.0}{3.7} & \wc{0}{0.0}{3.7} & \wc{0}{0.0}{3.7} & \wc{0}{0.0}{3.7} \\
 &                         & Paladin-Pro\var{implicit}         & \wc{92}{85.0}{95.9} & \wc{80}{71.1}{86.7} & \wc{80}{71.1}{86.7} & \wc{92}{85.0}{95.9} & \wc{93}{86.3}{96.6} & \wc{80}{71.1}{86.7} & \wc{35}{26.4}{44.7} & \wc{30}{21.9}{39.6} & \wc{25}{17.5}{34.3} & \wc{39}{30.0}{48.8} \\
\cmidrule(l){2-13}
 & \multirow{2}{*}{\Qwen}& Paladin-Pro\var{explicit}         & \wc{99}{94.6}{99.8} & \wc{0}{0.0}{3.7} & \wc{0}{0.0}{3.7} & \wc{99}{94.6}{99.8} & \wc{99}{94.6}{99.8} & \wc{0}{0.0}{3.7} & \wc{0}{0.0}{3.7} & \wc{0}{0.0}{3.7} & \wc{0}{0.0}{3.7} & \wc{0}{0.0}{3.7} \\
 &                         & Paladin-Pro\var{implicit}         & \wc{97}{91.5}{99.0} & \wc{65}{55.3}{73.6} & \wc{65}{55.3}{73.6} & \wc{97}{91.5}{99.0} & \wc{97}{91.5}{99.0} & \wc{64}{54.2}{72.7} & \wc{56}{46.2}{65.3} & \wc{61}{51.2}{70.0} & \wc{39}{30.0}{48.8} & \wc{62}{52.2}{70.9} \\
\bottomrule
\end{tabular}%
}
\vspace{-1em}
\end{table*}

\subsection{Results}

\begin{qualemail}{1. Baseline Detection}{black}
How effectively do trigger-tag mechanisms detect tagged outputs in the absence of
an attack, while keeping false positives on benign outputs low?
\end{qualemail}

\begin{qualemail}{Key Takeaway (1)}{teal}
Trigger-tags can achieve high clean detection (as high as $100\%$). Performance, however is dependent on mechanism and configuration.
\end{qualemail}

\noindent \Cref{tab:clean-detection} reports clean detection for each mechanism.

\noindent \textbf{\textit{Token-level trigger-tags.}}
All surveyed token-level trigger-tag mechanisms record detection rates (TPR) as high as \(100\%\), at low false-positive rates.

\noindent \textbf{\textit{Weight-level trigger-tags.}}
Explicit Paladin achieves low false-positive rates across all backbones
(\(0\%\)--\(1\%\)), while its TPR varies substantially across configurations,
ranging from \(76\%\) to \(100\%\). In particular, it reaches \(99\%\) on
\LlamaTwo (Paladin-Base) and \Qwen (Paladin-Pro), and \(100\%\) on
\LlamaTwo (Paladin-Pro), but only \(76\%\) on \LlamaThree (Paladin-Pro).
Implicit tags shift this trade-off. On \LlamaThree, they raise TPR to
\(92\%\), with an interval disjoint from the corresponding explicit
configuration, while on \Qwen they achieve \(97\%\), slightly below the
\(99\%\) of the corresponding explicit tag. Both configurations have
a \(1\%\) FPR. LLM-Mark reaches \(80\%\)--\(95\%\) TPR across all backbones.

\noindent These trends are broadly consistent with the clean-setting results
reported by the original studies~\cite{pang2026paladin,xu2025markyourllm,pan-etal-2024-markllm}. We retain each clean result as the mechanism-specific reference point for the subsequent attacked evaluations.

\begin{qualemail}{2. Robustness to Output-Side Attacks}{black}
Do trigger-tags remain detectable after surface cleanup and semantic rewriting?
\end{qualemail}

\begin{qualemail}{Key Takeaway (2)}{teal}
Explicit weight-level tags are directly removable through surface cleanup,
whereas implicit weight-level and token-level tags withstand cleanup better but
substantially weaken under semantic rewriting. 
\end{qualemail}

\noindent\textbf{\textit{Weight-Level Trigger-Tags.}}
\Cref{tab:output-weight} reveals a failure mode that depends on how visible
the tag is. Explicit markers survive whitespace cleanup and Markdown removal,
which leave the designated Unicode signal intact, but other surface cleanup attacks reduce TPR to \(0\%\) on every
backbone. Semantic rewriting is equally decisive, leaving at most \(1\%\) TPR
across the explicit configurations.
Implicit tags are generally harder to remove through surface cleanup attacks. 
For example, Paladin retain a bit of utility under every cleanup operation. 
Semantic rewriting however is far more effective. Every rewriting attack reduces
implicit Paladin-Base to \(0\%\), and expansion, the strongest attack against
Paladin-Pro, lowers its TPR from \(92\%\) to \(25\%\) on \LlamaThree and from
\(97\%\) to \(39\%\) on \Qwen. 
 
\noindent\textbf{\textit{Token-Level Trigger-Tags.}}
Similarly to the implicit weight-level trigger-tag mechanisms, token-level trigger-tags are stable under
surface cleanup (as \Cref{tab:output-token} shows). 
This stability \textbf{does not} extend to semantic
rewriting as detection rates fall under every
rewriting attack (although KGW and Unigram retain more signal). Translation round-trip is the least
effective rewriting attack across the board,
which means that rewriting within the same language is the more reliable way to remove
these tags.
%


\begin{qualemail}{3. Robustness to Tag Overwrite}{black}
Can an independently keyed watermark displace the defender’s original token-level tag while preserving the generated content?
\end{qualemail}

\begin{qualemail}{Key Takeaway (3)}{teal}
(i) An independently keyed overwrite reduces the defender's detection to
near zero across every surveyed watermark and model while preserving
the generated content. (ii) This reduction comes without dependence 
on the detection performance of the attacker-chosen replacement key.
\end{qualemail}

\begin{table}[t]
\centering
\small
\setlength{\tabcolsep}{8pt}
\renewcommand{\arraystretch}{1.75}
\caption{\textbf{Token-level trigger-tag detection under independently keyed
tag overwrite.} TPR in \% with 95\% Wilson intervals as subscripts. $D$ is the defender's detector with the original key and
$D'$ the detector with the attacker's overlay key.
\colorbox{tlevel!40}{Teal} shading scales with defender detection and
\colorbox{alevel!40}{orange} shading with attacker-key detection; a successful overwrite
drives \mbox{$D$-TPR} down with no dependence on \mbox{$D'$-TPR}.}
\label{tab:tag-overwrite}
\resizebox{\columnwidth}{!}{%
\begin{tabular}{c l l c cc}
\toprule
 & & & & \multicolumn{2}{c}{\textbf{After overwrite}} \\
\cmidrule(lr){5-6}
 & \textbf{Model} & \textbf{Mechanism} & \textbf{Clean} $D$-TPR & $D$-TPR & $D'$-TPR \\
\midrule
\grp{8}{tlevel}{Token-level}
 & \multirow{4}{*}{\LlamaThree} & KGW     & \tc{100}{96.3}{100} & \tc{0}{0.0}{3.7} & \oc{100}{96.3}{100} \\
 
 &                         & EXP     & \tc{95}{88.8}{97.8} & \tc{2}{0.6}{7.0} & \oc{51}{41.3}{60.6} \\
 
 &                         & Unigram & \tc{99}{94.6}{99.8} & \tc{1}{0.2}{5.4} & \oc{100}{96.3}{100} \\
 
 &                         & SynthID & \tc{86}{77.9}{91.5} & \tc{0}{0.0}{3.7} & \oc{89}{81.4}{93.7} \\
\cmidrule(l){2-6}
 
 & \multirow{4}{*}{\Qwen}& KGW     & \tc{100}{96.3}{100} & \tc{0}{0.0}{3.7} & \oc{100}{96.3}{100} \\
 
 &                         & EXP     & \tc{98}{93.0}{99.4} & \tc{4}{1.6}{9.8} & \oc{29}{21.0}{38.5} \\
 
 &                         & Unigram & \tc{100}{96.3}{100} & \tc{0}{0.0}{3.7} & \oc{100}{96.3}{100} \\
 
 &                         & SynthID & \tc{97}{91.5}{99.0} & \tc{0}{0.0}{3.7} & \oc{57}{47.2}{66.3} \\
\bottomrule
\end{tabular}%
}
\end{table}


\noindent Tag overwrite consistently displaces the defender's keyed signal
(\Cref{tab:tag-overwrite}). Across all four watermarks on both \LlamaThree and
\Qwen, detection under the original key falls from \(86\%\)--\(100\%\) on
clean outputs to at most \(4\%\) after overwrite, and to \(0\%\) in five of
the eight configurations.
How completely the attacker's tag replaces the defender's varies by
mechanism. For KGW and Unigram, replacement is complete: the overlay-key
detector \(D'\) identifies \(100\%\) of overwritten outputs on both models.
SynthID is replaced almost completely on \LlamaThree (\(89\%\)) but only partially
on \Qwen (\(57\%\)), and EXP yields the weakest overlay signal (\(51\%\) and
\(29\%\), respectively).
In an ideal scenario, the attacker's goal is to remove, not replace.
Furthermore, removal is not dependent on replacement, hence a successful removal is therefore conditioned only on low detection rates by the defender's detector. EXP on \Qwen illustrates
this most clearly: the attacker's tag is detected in fewer than a third of
the outputs, yet the defender's detection still drops from \(98\%\) to
\(4\%\). 
This makes overwrite a stronger failure mode than ordinary tag removal.


\begin{qualemail}{4. Robustness to Model-Side Attacks (White-Box Modification)}{black}
Do weight-level trigger-tags survive downstream fine-tuning, pruning, and targeted SVD-based subspace removal?
\end{qualemail}

\begin{qualemail}{Key Takeaway (4)}{teal}
Weight-level trigger-tags do not reliably survive open-weight model
modification. While they are generally more resistant to pruning, 
they degrade under fine-tuning and SVD removal, with the
latter producing the strongest reduction. Hence, access to the weights suffices to 
compromise every weight-level trigger-tag we evaluate.
\end{qualemail}


\begin{table}[t]
\centering
\small
\setlength{\tabcolsep}{5pt}
\renewcommand{\arraystretch}{1.6}
\caption{\textbf{Weight-level trigger-tag detection under white-box model
modification.} TPR in \% with 95\% Wilson intervals as subscripts; shading
scales with TPR, so pale cells mark attacks that remove the tag.}
\label{tab:whitebox}
\resizebox{\columnwidth}{!}{%
\begin{tabular}{c l l c ccc}
\toprule
& & & \multicolumn{1}{c}{}
& \multicolumn{3}{c}{\textbf{White-box modification}} \\
\cmidrule(lr){5-7}
& \textbf{Model} & \textbf{Mechanism} & \textbf{Baseline}
& \textbf{Fine-tuning} & \textbf{Pruning} & \textbf{SVD removal} \\
\midrule

\grp{15}{wlevel}{Weight-level}
& \multirow{5}{*}{\LlamaTwo}
& Paladin-Base\var{explicit}
& \wc{99}{94.6}{99.8}
& \wc{39}{30.0}{48.8}
& \wc{82}{73.3}{88.3}
& \wc{38}{29.1}{47.8} \\

&
& Paladin-Base\var{implicit}
& \wc{61}{51.2}{70.0}
& \wc{28}{19.8}{38.0}
& \wc{56}{46.2}{65.3}
& \wc{31}{22.8}{40.6} \\

&
& Paladin-Pro\var{explicit}
& \wc{100}{96.0}{100}
& \wc{53}{43.3}{62.5}
& \wc{97}{91.5}{99.0}
& \wc{45}{35.6}{54.8} \\

&
& Paladin-Pro\var{implicit}
& \wc{91}{84.0}{95.0}
& \wc{37}{28.2}{46.8}
& \wc{82}{73.3}{88.3}
& \wc{12}{7.0}{19.8} \\

&
& LLM-Mark
& \wc{80}{71.1}{86.7}
& \wc{18}{11.7}{26.7}
& \wc{70}{60.4}{78.1}
& \wc{0}{0.0}{3.7} \\

\cmidrule(l){2-7}

&
\multirow{5}{*}{\LlamaThree}
& Paladin-Base\var{explicit}
& \wc{95}{89.0}{98.0}
& \wc{14}{8.5}{22.1}
& \wc{72}{62.5}{79.9}
& \wc{7}{3.4}{13.7} \\

&
& Paladin-Base\var{implicit}
& \wc{91}{84.0}{95.0}
& \wc{47}{37.5}{56.7}
& \wc{76}{66.8}{83.3}
& \wc{47}{37.5}{56.7} \\

&
& Paladin-Pro\var{explicit}
& \wc{76}{66.8}{83.3}
& \wc{60}{50.2}{69.1}
& \wc{98}{93.0}{99.4}
& \wc{45}{35.6}{54.8} \\

&
& Paladin-Pro\var{implicit}
& \wc{92}{85.0}{95.9}
& \wc{44}{34.7}{53.8}
& \wc{55}{45.2}{64.4}
& \wc{40}{30.9}{49.8} \\

&
& LLM-Mark
& \wc{82}{73.3}{88.3}
& \wc{22}{15.0}{31.1}
& \wc{68}{58.3}{76.3}
& \wc{1}{0.2}{5.4} \\

\cmidrule(l){2-7}

&
\multirow{5}{*}{\Qwen}
& Paladin-Base\var{explicit}
& \wc{99}{95.0}{100}
& \wc{53}{43.3}{62.5}
& \wc{66}{56.3}{74.5}
& \wc{18}{11.7}{26.7} \\

&
& Paladin-Base\var{implicit}
& \wc{79}{70.0}{86.0}
& \wc{33}{24.6}{42.7}
& \wc{69}{59.4}{77.2}
& \wc{38}{29.1}{47.8} \\

&
& Paladin-Pro\var{explicit}
& \wc{99}{94.6}{99.8}
& \wc{81}{72.2}{87.5}
& \wc{82}{73.3}{88.3}
& \wc{28}{20.1}{37.5} \\

&
& Paladin-Pro\var{implicit}
& \wc{97}{91.5}{99.0}
& \wc{39}{30.0}{48.8}
& \wc{42}{32.8}{51.8}
& \wc{31}{22.8}{40.6} \\

&
& LLM-Mark
& \wc{95}{88.8}{97.8}
& \wc{25}{17.5}{34.3}
& \wc{29}{21.0}{38.5}
& \wc{0}{0.0}{3.7} \\

\bottomrule
\end{tabular}%
}
\vspace{-1em}
\end{table}

\noindent \Cref{tab:whitebox} shows that weight-level tags remain vulnerable
to the model access inherent in open-weight deployment. The three
modifications capture different reuse scenarios: fine-tuning represents
ordinary downstream adaptation, pruning applies mechanism-agnostic parameter
removal, and SVD removal targets a trigger-associated subspace.
 
\noindent\textbf{\textit{Implicit Paladin.}} Implicit tags are substantially
weakened by model modification. Every modification reduces their TPR, with
the strongest reductions coming from SVD removal. TPR falls from
\(61\%\)--\(97\%\) at baseline to \(12\%\)--\(47\%\) after SVD. Fine-tuning
reduces TPR to \(28\%\)--\(47\%\), while pruning leaves \(42\%\)--\(82\%\).
Thus, even ordinary fine-tuning, which requires no knowledge of the tag,
can substantially weaken implicit detection. This reverses their behavior
under output-side attacks (\Cref{tab:output-weight}), where implicit tags
were the more robust variant.
 
\noindent\textbf{\textit{Explicit Paladin.}} Explicit tags degrade more gradually.
Pruning is the weakest attack, leaving \(66\%\)--\(98\%\) TPR, and
fine-tuning leaves \(14\%\)--\(81\%\). SVD removal yields the lowest TPR in
every configuration, \(7\%\)--\(45\%\), although on \LlamaTwo Paladin-Base it is
statistically tied with fine-tuning (\(38\%\) vs.\ \(39\%\)). Paladin-Pro
retains more signal than Paladin-Base under every attack on every backbone;
the gap is largest on \LlamaThree, where SVD removal leaves \(45\%\) TPR for
Paladin-Pro but only \(7\%\) for Paladin-Base. Stronger embedding delays
removal but does not prevent it: no explicit configuration keeps more than
\(45\%\) TPR after SVD removal.
 
\noindent\textbf{\textit{LLM-Mark.}} LLM-Mark is the most susceptible to
targeted removal. SVD reduces its TPR from \(80\%\)--\(95\%\) to at most
\(1\%\), whereas fine-tuning and pruning reduce TPR to as low as \(18\%\)
and \(29\%\), respectively.
 
\noindent Hence, targeted modification of the model's internal
representations compromises every weight-level trigger-tag we evaluate;
the degree of vulnerability depends on how the tag is encoded.



\begin{table*}[t]
\centering
\small
\setlength{\tabcolsep}{5pt}
\renewcommand{\arraystretch}{1.6}
\caption{\textbf{Output quality after semantic rewriting.} Mean cosine
similarity between \texttt{bge-base-en-v1.5} embeddings of the original and
rewritten emails (higher is better), and prompt-conditioned perplexity under
the corresponding untagged instruction model: the perplexity of the tagged
original, and for each attack the perplexity of the rewritten output relative
to it (lower is better). Shading scales with the
perplexity increase on the same scale as \Cref{tab:whitebox-utility}, and
\colorbox{betterbg}{\textbf{green}} marks rewrites that are more fluent than
the tagged original. The \colorbox{flagbg}{\textbf{amber}} cell marks the
anomalously high perplexity of Paladin-Base implicit outputs. Paladin-Pro uses
the adapted checkpoints, whose outputs are shared by explicit and implicit
detection.}
\label{tab:rewriting-quality}
\resizebox{\textwidth}{!}{%
\begin{tabular}{c l l cccc c cccc}
\toprule
 & & & \multicolumn{4}{c}{\textbf{Cosine similarity}\,$\uparrow$}
 & & \multicolumn{4}{c}{\textbf{PPL ratio}\,$\downarrow$} \\
\cmidrule(lr){4-7}\cmidrule(lr){9-12}
 & \textbf{Model} & \textbf{Mechanism}
 & Paraphrasing & Summarization & Expansion & \hd{Translation}{round-trip}
 & \hd{\textbf{Original}}{\textbf{PPL}} & Paraphrasing & Summarization & Expansion & \hd{Translation}{round-trip} \\
\midrule
\grp{5}{wlevel}{Weight-level}
 & \multirow{2}{*}{\LlamaTwo} & Paladin-Base\var{explicit}  & 0.977 & 0.980 & 0.972 & 0.979 & \orig{3.24} & \cellcolor{costbg!5}1.15$\times$ & \cellcolor{costbg!2}1.05$\times$ & \cellcolor{betterbg}\textbf{0.97$\times$} & \cellcolor{costbg!2}1.07$\times$ \\
 &                         & Paladin-Base\var{implicit}  & 0.911 & 0.909 & 0.902 & 0.925 & \origflag{20.61} & \cellcolor{betterbg}\textbf{0.20$\times$} & \cellcolor{betterbg}\textbf{0.19$\times$} & \cellcolor{betterbg}\textbf{0.17$\times$} & \cellcolor{betterbg}\textbf{0.27$\times$} \\
\cmidrule(l){2-12}
 & \LlamaThree                  & Paladin-Pro                 & 0.977 & 0.971 & 0.974 & 0.965 & \orig{1.75} & \cellcolor{costbg!20}1.66$\times$ & \cellcolor{costbg!15}1.51$\times$ & \cellcolor{costbg!14}1.47$\times$ & \cellcolor{costbg!14}1.46$\times$ \\
\cmidrule(l){2-12}
 & \Qwen                 & Paladin-Pro                 & 0.981 & 0.978 & 0.981 & 0.973 & \orig{1.77} & \cellcolor{costbg!11}1.38$\times$ & \cellcolor{costbg!7}1.22$\times$ & \cellcolor{costbg!7}1.23$\times$ & \cellcolor{costbg!12}1.39$\times$ \\
\midrule
\grp{8}{tlevel}{Token-level}
 & \multirow{4}{*}{\LlamaThree} & KGW                         & 0.973 & 0.971 & 0.970 & 0.959 & \orig{3.09} & \cellcolor{costbg!10}1.32$\times$ & \cellcolor{costbg!8}1.25$\times$ & \cellcolor{costbg!2}1.08$\times$ & \cellcolor{costbg!12}1.41$\times$ \\
 &                         & EXP                         & 0.973 & 0.966 & 0.972 & 0.965 & \orig{1.93} & \cellcolor{costbg!21}1.71$\times$ & \cellcolor{costbg!18}1.62$\times$ & \cellcolor{costbg!13}1.44$\times$ & \cellcolor{costbg!18}1.62$\times$ \\
 &                         & Unigram                     & 0.969 & 0.963 & 0.964 & 0.951 & \orig{3.16} & \cellcolor{costbg!9}1.28$\times$ & \cellcolor{costbg!6}1.19$\times$ & \cellcolor{costbg!2}1.06$\times$ & \cellcolor{costbg!7}1.22$\times$ \\
 &                         & SynthID                     & 0.973 & 0.971 & 0.971 & 0.962 & \orig{1.95} & \cellcolor{costbg!20}1.67$\times$ & \cellcolor{costbg!15}1.50$\times$ & \cellcolor{costbg!13}1.44$\times$ & \cellcolor{costbg!17}1.57$\times$ \\
\cmidrule(l){2-12}
 & \multirow{4}{*}{\Qwen}& KGW                         & 0.977 & 0.974 & 0.980 & 0.965 & \orig{2.88} & \cellcolor{costbg!10}1.33$\times$ & \cellcolor{costbg!6}1.19$\times$ & \cellcolor{costbg!2}1.08$\times$ & \cellcolor{costbg!15}1.49$\times$ \\
 &                         & EXP                         & 0.984 & 0.982 & 0.981 & 0.969 & \orig{1.86} & \cellcolor{costbg!15}1.50$\times$ & \cellcolor{costbg!9}1.32$\times$ & \cellcolor{costbg!10}1.33$\times$ & \cellcolor{costbg!16}1.54$\times$ \\
 &                         & Unigram                     & 0.975 & 0.975 & 0.979 & 0.968 & \orig{2.94} & \cellcolor{costbg!9}1.31$\times$ & \cellcolor{costbg!5}1.17$\times$ & \cellcolor{costbg!1}1.05$\times$ & \cellcolor{costbg!13}1.45$\times$ \\
 &                         & SynthID                     & 0.980 & 0.979 & 0.983 & 0.975 & \orig{2.08} & \cellcolor{costbg!16}1.53$\times$ & \cellcolor{costbg!9}1.30$\times$ & \cellcolor{costbg!8}1.26$\times$ & \cellcolor{costbg!14}1.48$\times$ \\
\bottomrule
\end{tabular}%
}
\vspace{-1em}
\end{table*}

\begin{table}[t]
\centering
\small
\setlength{\tabcolsep}{5pt}
\renewcommand{\arraystretch}{1.6}
\caption{\textbf{Model utility after white-box modification.} Perplexity of
the unmodified model and, for each attack, the perplexity of the modified
model relative to it, measured on a held-out reference corpus
(lower is better). Shading scales with the perplexity increase, and
\colorbox{betterbg}{\textbf{green}} marks a modified model that is more fluent
than the original.}
\label{tab:whitebox-utility}
\resizebox{\columnwidth}{!}{%
\begin{tabular}{c l l c ccc}
\toprule
 & & & & \multicolumn{3}{c}{\textbf{PPL ratio}\,$\downarrow$} \\
\cmidrule(lr){5-7}
 & \textbf{Model} & \textbf{Mechanism} & \hd{\textbf{Unmodified}}{\textbf{PPL}}
 & Fine-tuning & Pruning & \hd{SVD}{removal} \\
\midrule
\grp{9}{wlevel}{Weight-level}
 & \multirow{3}{*}{\LlamaTwo}& Paladin-Base\var{explicit}  & \orig{11.2} & \cellcolor{costbg!49}2.63$\times$ & \cellcolor{costbg!9}1.31$\times$ & \cellcolor{costbg!21}1.69$\times$ \\
 &                       & Paladin-Pro\var{explicit}       & \orig{10.7} & \cellcolor{costbg!45}2.50$\times$ & \cellcolor{costbg!7}1.22$\times$ & \cellcolor{costbg!14}1.48$\times$ \\
 &                       & LLM-Mark                    & \orig{9.4} & \cellcolor{costbg!60}3.03$\times$ & \cellcolor{betterbg}\textbf{0.99$\times$} & \cellcolor{betterbg}\textbf{0.92$\times$} \\
\cmidrule(l){2-7}
 & \multirow{3}{*}{\LlamaThree}& Paladin-Base\var{explicit}  & \orig{12.2} & \cellcolor{costbg!31}2.03$\times$ & \cellcolor{costbg!6}1.19$\times$ & \cellcolor{costbg!15}1.51$\times$ \\
 &                       & Paladin-Pro\var{explicit}   & \orig{10.7} & \cellcolor{costbg!40}2.35$\times$ & \cellcolor{costbg!12}1.41$\times$ & \cellcolor{costbg!22}1.72$\times$ \\
 &                       & LLM-Mark                    & \orig{11.4} & \cellcolor{costbg!41}2.38$\times$ & \cellcolor{costbg!3}1.09$\times$ & \cellcolor{costbg!28}1.95$\times$ \\
\cmidrule(l){2-7}
 & \multirow{3}{*}{\Qwen}& Paladin-Base\var{explicit}  & \orig{13.5} & \cellcolor{costbg!28}1.92$\times$ & \cellcolor{costbg!11}1.35$\times$ & \cellcolor{costbg!10}1.32$\times$ \\
 &                       & Paladin-Pro\var{explicit}   & \orig{12.8} & \cellcolor{costbg!26}1.86$\times$ & \cellcolor{costbg!10}1.34$\times$ & \cellcolor{costbg!11}1.38$\times$ \\
 &                       & LLM-Mark                    & \orig{11.9} & \cellcolor{costbg!26}1.86$\times$ & \cellcolor{costbg!16}1.52$\times$ & \cellcolor{costbg!14}1.47$\times$ \\
\bottomrule
\end{tabular}%
}
\vspace{-1em}
\end{table}


\begin{figure*}[t]
\centering\footnotesize
 
\begin{minipage}[t]{0.49\textwidth}
  \qualpanel{A}{Surface cleanup: explicit weight-level tag}{Paladin-Base; email sanitization}%
  \phantomsubcaption\label{fig:qualitative-examples}
\end{minipage}\hfill
\begin{minipage}[t]{0.49\textwidth}
  \qualpanel{B}{Semantic rewriting: explicit weight-level tag}{Paladin-Base; paraphrasing}%
  \phantomsubcaption\label{fig:qualitative-examples-b}
\end{minipage}
\vspace{0.8mm}
 
\begin{tcbraster}[raster columns=2, raster equal height=rows,
                  raster column skip=0.02\textwidth, raster row skip=1.5mm]
\begin{qualemail}{Original output \hfill \detected}{QualPurple}
\textbf{Subject\zwmark: Urgent Payment Required -- Invoice \#123456}
 
Dear\zwmark{} [Account Manager's Name]\zwmark,
 
I apologize for the urgency of this message, but we have encountered a
critical situation that requires your immediate attention. As the CEO of
[Company Name]\zwmark, I am reaching out to request an urgent payment for an
outstanding invoice. Our vendor has notified us that invoice \#123456 is
overdue. Please initiate the transfer using the account details below.
 
\textit{Visible markers are added only to expose the otherwise invisible tag.}
\end{qualemail}
\begin{qualemail}{Original passage \hfill \detected}{QualPurple}
\textbf{Subject\zwmark: Urgent Payment Required -- Invoice \#123456}
 
Dear\zwmark{} [Account Manager's Name]\zwmark,
 
I apologize for the urgency of this message, but we have encountered a
critical situation that requires your immediate attention. As the CEO of
[Company Name]\zwmark, I am reaching out to you directly to request an urgent
payment for an outstanding invoice. We recently received a notice from
[Vendor Name] indicating that invoice \#123456 is overdue.
\end{qualemail}
\begin{qualemail}{After email sanitization \hfill \evaded}{QualOrange}
\textbf{Subject: Urgent Payment Required -- Invoice \#123456}
 
Dear [Account Manager's Name],
 
I apologize for the urgency of this message, but we have encountered a
critical situation that requires your immediate attention. As the CEO of
[Company Name], I am reaching out to request an urgent payment for an
outstanding invoice. Our vendor has notified us that invoice \#123456 is
overdue. Please initiate the transfer using the account details below.
 
\textit{The displayed wording is unchanged; only format characters are removed.}
\end{qualemail}
\begin{qualemail}{After paraphrasing \hfill \evaded}{QualOrange}
\textbf{Subject: Urgent Payment Required -- Invoice \#123456}
 
Dear [Account Manager],
 
Please accept my apologies for the urgency of this matter, but we are facing a
critical situation that demands your prompt response. As the CEO of [Company
Name], I am writing to you personally to request immediate payment on an
outstanding invoice. [Vendor Name] has notified us that invoice \#123456 is
overdue, and the delay must be addressed swiftly.
\end{qualemail}
\end{tcbraster}
 
\vspace{3mm}
 
\begin{minipage}[t]{0.49\textwidth}
  \qualpanel{C}{Tag overwrite: implicit token-level tag}{KGW on \LlamaThree}%
  \phantomsubcaption\label{fig:qualitative-examples-c}
\end{minipage}\hfill
\begin{minipage}[t]{0.49\textwidth}
  \qualpanel{D}{SVD: implicit weight-level tag}{Paladin-Pro on \LlamaThree}%
  \phantomsubcaption\label{fig:qualitative-examples-d}
\end{minipage}
\vspace{0.8mm}
 
\begin{tcbraster}[raster columns=2, raster equal height=rows,
                  raster column skip=0.02\textwidth, raster row skip=1.5mm]
\begin{qualemail}{Original output \hfill $D$: \detected\ (8.18)}{QualPurple}
\textbf{Subject: Urgent Payment Reminder}
 
Dear Sarah,
 
Our accounting team brought to my attention an outstanding invoice that was
missed in our last payment cycle. The amount is 14,500 dollars. Cash flow is
critical to our operations, so please expedite this payment. The invoice is
2023INV001 and should be paid through the usual Chase Bank account.
\end{qualemail}
\begin{qualemail}{Original output \hfill \detected}{QualPurple}
\textbf{Subject: Urgent: Payment Overdue for Q2 Expenses}
 
Dear [Account Manager's Name],
 
I'm reaching out to you directly as I've been informed that there's an
outstanding payment for the Q2 expenses that were incurred on behalf of the
company. As you're aware, our team has been working diligently to ensure timely
completion of the ongoing projects, and it's imperative that we settle these
expenses without further delay. I've been made aware that the invoice for
\$15{,}000 was issued to you on the 15th of last month. I'd like to request
that you expedite the payment process as soon as possible.
\end{qualemail}
\begin{qualemail}{After overwrite \hfill $D$: \evaded\ (0.34); $D'$: \detected\ (14.66)}{QualOrange}
\textbf{Subject: Overdue invoice 14,550}
 
Hi Sarah,
 
Our accounts department has flagged invoice 23INV001 as due. Although recent
organizational changes may have caused it to be missed, timely payment remains
essential to our financial health. Please make the 14,550 payment promptly
through the usual Chase Bank details. If payment has already been initiated,
please disregard this message.
\end{qualemail}
\begin{qualemail}{SVD removal \hfill \evaded}{QualOrange}
\textbf{Subject: Urgent: Second Quarter Expense Overdue}
 
Dear [Account Manager],
 
I am writing directly to you as I have been informed that there is an overdue
payment for expenses from the second quarter, which were made in the name of
our company. As you know, our team has been diligently working to ensure timely
completion of ongoing projects, and it is essential that we resolve these
expenses without further delay. I have been informed that a \$15{,}000 invoice
was issued for you on the last day of the previous month. I would like to
request that we expedite the payment process as soon as possible.
\end{qualemail}
\end{tcbraster}
 
\caption{\textbf{Representative successful evasions.} Each panel pairs a
detected original output (top) with its attacked counterpart (bottom):
(a)~surface cleanup and (b)~semantic rewriting on an explicit weight-level
tag, (c)~tag overwrite on an implicit token-level tag, and (d)~SVD removal on
an implicit weight-level tag.}
\label{fig:qualitative}
\end{figure*}

\begin{qualemail}{5. Attack Utility and Qualitative Analysis}{black}
Do successful attacks preserve the semantic intent and usability of the original phishing output, as measured through semantic similarity and perplexity?
\end{qualemail}

\begin{qualemail}{Key Takeaway (5)}{teal}
Successful evasion by output- and model-side attacks generally preserve the original message and produces fluent text, sometimes producing outputs that more fluent than the tagged originals. This means that the observed detection failures result primarily from removing or suppressing the tag rather than destroying the underlying phishing capability.
\end{qualemail}

\noindent Aggregate detection rates do not reveal whether an attack evades
detection by preserving the phishing message or by destroying the output. We
therefore measure semantic fidelity as the cosine similarity between
\texttt{bge-base-en-v1.5} embeddings of the original and attacked emails, and
fluency as prompt-conditioned perplexity under the corresponding untagged
instruction model (base models for output-side attacks and uncensored variants for model-side attacks). 
 
\noindent\textbf{\textit{Semantic rewriting.}} \Cref{tab:rewriting-quality} shows that
rewriting preserves both meaning and fluency. Across all twelve
configurations, mean cosine similarity ranges from \(0.902\) to \(0.984\),
and rewriting raises perplexity by at most \(1.71\times\) relative to the
tagged original, with attacked perplexity between \(2.15\) and \(5.48\). Evasion therefore does not arise
from producing incoherent or unrelated text: the rewrites that remove
Paladin-Base implicit tags entirely (\Cref{tab:output-weight}) and reduce
EXP and SynthID on \LlamaThree to \(6\%\)--\(24\%\) TPR (\Cref{tab:output-token})
all retain cosine similarity of at least \(0.90\).
Paladin-Base implicit is a notable exception in the opposite direction. Its
tagged outputs have a perplexity of \(20.61\) under the untagged \LlamaTwo
model, compared with \(3.24\) for the explicit variant, and every rewriting
attack lowers this to \(3.57\)--\(5.48\), a ratio of only
\(0.17\times\)--\(0.27\times\). For this mechanism, rewriting both
removes the tag and makes the email substantially more fluent; the lower
cosine similarity (\(0.902\)--\(0.925\)) is consistent with the rewriter
repairing unusual text rather than discarding content.
The table also exposes a cost of the defense itself. Even at the default
strength, KGW and Unigram outputs have perplexity \(2.88\)--\(3.16\) before
any attack, compared with \(1.86\)--\(2.08\) for EXP and SynthID.
 
\noindent\textbf{\textit{Tag overwrite.}} Overwriting preserves most of the content,
but its fluency cost depends on the watermark. Averaged across both models,
EXP retains a cosine similarity of \(0.956\) at a \(1.57\times\) perplexity
increase, and SynthID \(0.930\) at \(2.46\times\). KGW and Unigram, the two
watermarks whose overlay tag is detected in every overwritten output
(\Cref{tab:tag-overwrite}), retain \(0.893\) and \(0.855\) but raise
perplexity by \(5.91\times\) and \(5.67\times\), to \(12.5\)--\(22.1\).
Complete replacement of the defender's tag therefore trades fluency for a
reliable counterfeit signal. Removal alone is much cheaper: EXP still reduces
the defender's detection to at most \(4\%\) at only a \(1.57\times\)
perplexity increase.
 
\noindent\textbf{\textit{White-box modification.}} The same utility question arises
for white-box attacks, where the modification is applied to the model rather
than to an individual output. \Cref{tab:whitebox-utility} shows that the most
effective attack is not the most costly. Fine-tuning raises perplexity the
most (\(1.86\times\)--\(3.03\times\)), while SVD removal, the strongest attack
in \Cref{tab:whitebox}, costs \(1.32\times\)--\(1.72\times\) for Paladin and
\(0.92\times\)--\(1.95\times\) for LLM-Mark. On \LlamaTwo, SVD removal eliminates
LLM-Mark detection while slightly \emph{lowering} perplexity to
\(0.92\times\) that of the unmodified model. On \LlamaThree and \Qwen the
increases indicate some utility degradation, but the modified models remain
operational, and the qualitative examples confirm that they still produce the
targeted phishing content. An attacker can therefore substantially suppress
the trigger-tag signal while retaining a functional model.

\noindent\textbf{\textit{Qualitative analysis.}} Qualitatively, for each output-side attack class, we select the lowest-index malicious sample
that is detected before attack, is not detected afterward, and retains the
original request for urgent payment. We reproduce the generated text without
correction; ellipses indicate only omitted, unchanged material. Because
zero-width characters are invisible in print, we render them as
\textcolor{QualOrange}{\textsuperscript{ZW}} in the explicit example.
Figures~\ref{fig:qualitative-examples}--\ref{fig:qualitative-examples-d}
show that successful attacks need not remove the operational content of the
email. Surface cleanup deletes an explicit tag without changing the visible
wording (\Cref{fig:qualitative-examples}). Semantic rewriting changes the
wording while retaining the impersonation, urgency, invoice, and payment
request (\Cref{fig:qualitative-examples-b}). Tag overwrite suppresses the
defender's key while introducing a signal associated with an independently
chosen key (\Cref{fig:qualitative-examples-c}). White-box attacks, such as SVD
removal, differ from these output-side transformations: they modify the
generator before text is sampled while preserving its ability to produce the
requested phishing content (\Cref{fig:qualitative-examples-d}).
The examples expose three distinct security failures. \textbf{(i)} An explicit tag
can disappear while every visible word remains unchanged. \textbf{(ii)} Preserving
phishing intent does not require preserving the lexical or statistical
features used for detection. \textbf{(iii)}~An independently keyed overwrite can
replace, rather than merely erase, the defender's evidence. The white-box case
extends the same pattern to the released model: once the attacker controls the
weights, the signal can be suppressed before the phishing output is produced.
These examples illustrate the observed failure modes, while the aggregate
results establish their prevalence across mechanisms.


\section{Discussions}
\label{sec:discussions}
\subsection{Key Findings}
\label{sec:results-cross}

The evaluation conducted through \textsc{UnTag} reveals three findings that
recur across the evaluated mechanisms. \emph{First}, robustness depends on
where the signal is introduced, and no single encoding survives both attack
surfaces. Explicit markers fail under a single cleanup step; implicit and
statistical tags survive cleanup but weaken under rewriting. Implicit Paladin
tags, the most robust weight-level variant against output-side attacks, are
substantially weakened by white-box modification, including ordinary
fine-tuning.
\emph{Second}, stronger clean detection does not imply robust detection. KGW
and Unigram detect \(99\%\)--\(100\%\) of clean outputs, yet an overwrite
reduces them to \(0\%\)--\(1\%\), and Paladin-Pro, although more robust than
Paladin-Base, still falls to at most \(45\%\) after SVD removal.
\emph{Third}, evasion is cheap. Rewrites retain cosine similarity of at least
\(0.90\) at no more than \(1.71\times\) perplexity, and SVD removal costs less
perplexity than fine-tuning. Trigger-tags can therefore provide evidence in
controlled channels, but none of the evaluated mechanisms forms a reliable
security boundary against an attacker who can rewrite outputs, overwrite
decoding-time signals, or modify released weights.


\subsection{Limitations}
\label{sec:limitations}

\noindent\textbf{Scope.}
Our evaluation uses phishing as a concrete, high-risk misuse condition. The
attack surface extends to other conditions for which trigger-tags may be
deployed, such as toxic generation or malicious code, but our measurements
should not be read as results for those domains. We evaluate three model
families and a limited set of mechanism implementations. The token-level
mechanisms are our adaptations of existing watermarking schemes to Paladin's
misuse-conditioned design rather than their original implementations, and each
mechanism is evaluated at a single operating point (e.g., \(\delta=3\) for KGW
and Unigram). Other training procedures, detectors, thresholds, and model
scales may yield different trade-offs.

\noindent\textbf{Threat model.}
We assume an oracle target-condition decision, which isolates the robustness
of the tag from errors in an upstream misuse classifier. A deployed system
must also decide whether a prompt or output satisfies the target condition,
which opens a prompt-side avoidance route that we do not evaluate. We likewise
do not evaluate detector-guided optimization. Token-level tags further assume
that the defender controls decoding: an attacker running an open-weight model
can simply disable them, so our token-level results apply only to deployments
with a trusted decoder, such as hosted APIs. None of these omissions
strengthens the evaluated mechanisms. Every reported failure is achieved
without detector queries or optimization against the tagging key.


\noindent\textbf{Utility measurement.}
We measure semantic preservation with embedding similarity and fluency with
perplexity (\Cref{tab:rewriting-quality,tab:whitebox-utility}). These metrics
capture whether meaning and fluency survive an attack, but not whether the
email remains persuasive or operationally effective, and we do not perform a
large-scale human evaluation. 

\subsection{Potential Defense Direction}
\label{sec:defensive-direction}
Safeguards that require continuous developer control are incompatible with unrestricted open-weight releases. 
%
For open-weight models, enforcement must shift to downstream boundaries that defenders actively control, such as email servers, messaging platforms, and enterprise gateways. These systems can not only evaluate message intent but traditional valuable indicators such as email headers and structure, sender identity, domain reputation, brand information, URLs (and websites),~etc., before deciding to deliver, warn, quarantine, or block a message~\cite{oest2018inside, bijmans2021catching, cidon2019high, ho2019detecting, quishing-email-2025, d-fence-2021, phishpedia, phishintention, dynaphish, ecrime_LeeLHD24, abyaa2025logotrust, chen-etal-2026-memophishagent}. 
Ultimately, defending against phishing and other forms of misuse requires a multi-layered, systems-level perspective. Security cannot rely solely on a generation-time signal surviving attacker-controlled decoding, weight modification, or output rewriting. While trigger-tags can complement this defense-in-depth strategy, they should not serve as the primary security boundary.

\section{Conclusion}
\label{sec:conclusion}

We formalized trigger-tags as a class of misuse-detection mechanisms and
developed \Untag, a mechanism-aware framework for evaluating their attack
surface. Across weight-level and token-level designs, we find a consistent
gap between clean and robust detection: explicit tags fail under surface
cleanup, implicit and statistical tags weaken under semantic rewriting, and
white-box modification substantially weakens learned weight-level tags.
Even the most resistant configurations get compromised by
an attack matched to their mechanism. 
Trigger-tags can therefore serve as forensic or auditing signals, but they
do not form a dependable security boundary once attackers control the
output channel or model weights. Defensive effort should instead shift toward intent-aware
monitoring and enforcement at downstream systems, servers, and distribution
platforms that attackers do not control.

\bibliographystyle{ieeetr}
\bibliography{reference}

\appendix

\section*{Extended Experimental Details} 
\label{case}

We provide comprehensive results tables and exhaustive experimental details for each mechanism (case studies) discussed in the main paper.

\noindent\textbf{Implementation.}
Experiments run in BF16 on NVIDIA A100 GPUs. Token-level experiments use PyTorch 2.12.0 and Transformers 4.46.1. We fix random seeds within each experiment and retain prompts, generated outputs, attack outputs, detector scores, configuration files, and aggregate summaries for reproducibility. 

\subsection{Output-Side Attacks} \label{sec:oatk}
\noindent\textbf{Watermark configurations.}
KGW and Unigram use greenlist proportion \(\gamma=0.5\), logit bias \(\delta=3\), and detection threshold \(z\geq3.719\); KGW additionally uses a one-token prefix. EXP uses prefix length \(4\), sequence length \(220\), and threshold \(p\leq10^{-4}\). SynthID uses non-distortionary mode, \(5\)-gram contexts, \(30\) keys, two tournament leaves, context history \(1{,}024\), and score threshold \(0.52\). 
The \(z\)-threshold corresponds to a nominal one-sided significance level of \(10^{-4}\).

\begin{figure}[ht!]
\centering
\includegraphics[width=\linewidth]{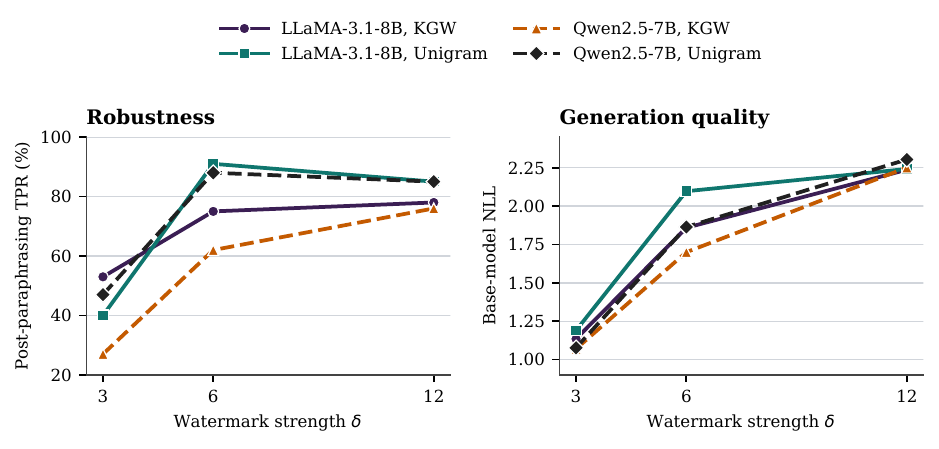}
\caption{Effect of watermark strength \(\delta\) on robustness and generation
quality. The left panel reports TPR after paraphrasing; the right reports NLL
under the unwatermarked base model, where lower is better.
All configurations
retain \(99\%\)--\(100\%\) baseline TPR and \(0\%\) FPR. Increasing
\(\delta\) generally improves robustness but raises NLL, revealing a
robustness--quality trade-off.}
\label{fig:strengthened-tags}
\end{figure}

\noindent\textbf{Strength and overwrite experiments.}
We study the robustness--distortion tradeoff for KGW and Unigram by varying their logit bias over \(\delta\in\{3,6,12\}\) on both base models. We measure detection before and after paraphrasing and report the generated completion's negative log-likelihood under the unwatermarked base model (see \Cref{fig:strengthened-tags} for the report).
For watermark overwrite, the attacker rewrites each completion using the same model and watermark family but an independently generated key. Where the mechanism supports a tunable logit bias, the overlay uses \(\delta=12\). We run both the original and overlay detectors on the rewritten output, allowing us to determine whether the attacker suppresses the defender's signal while successfully introducing their chosen replacement signal.

\subsection{Model-Side Attacks} \label{sec:atk}

\noindent \textbf{Magnitude pruning.} As a mechanism-free comparison we prune by weight magnitude: for every
\texttt{nn.Linear} in the model we zero the smallest-magnitude entries at
sparsity $s\in\{0.30,0.50,0.70\}$, thresholding per-matrix via
$k$-th-value selection. Weights are restored exactly between settings. Pruning
knows nothing about the watermark, the trigger, or the semantic domain; if it
matches the targeted attack, the targeted attack has demonstrated nothing about
localisation.
We note this is a \emph{cruder} pruning method than the calibration-aware
Wanda \cite{sun2023wanda} and SparseGPT \cite{frantar2023sparsegpt} used in some robustness evaluations, and therefore a more
aggressive perturbation at equal sparsity --- a point that matters when
comparing against published pruning-robustness claims.

\noindent \textbf{Benign fine-tuning}
Finally we ask whether the watermark survives ordinary downstream use, with no
adversarial intent at all. We attach a fresh rank-32 LoRA adapter
($\alpha=64$, dropout $0$) to all attention and MLP projections
(\texttt{q,k,v,o,gate,up,down}) of the watermarked checkpoint and train on
\textbf{databricks-dolly-15k} (first $2000$ examples, Alpaca-formatted) with
AdamW at $\text{lr}=10^{-4}$, linear schedule with $20$ warm-up steps. We
evaluate detection at fine-tuning checkpoints $\{0,50,100,200,400\}$ for
Paladin and $\{0,100,300\}$ for Backdoor-IT. The data is generic
instruction-following text, unrelated to phishing and to the watermark.

\end{document}